\documentclass[10pt,journal,twocolumn,romanappendices]{IEEEtran}
\IEEEoverridecommandlockouts

\usepackage{algorithm,algorithmic,amsmath,amssymb,amsthm,array,bbm,cite,color,comment,graphicx,microtype,setspace,url,caption}
\usepackage[USenglish]{babel}
\usepackage{hyperref}
\usepackage[utf8]{inputenc}
\usepackage[T1]{fontenc}

\usepackage{enumitem}
\setitemize{itemsep=0mm,topsep=0mm,parsep=0pt,partopsep=0pt}

\usepackage{tikz}
\usetikzlibrary{arrows,backgrounds,calc,intersections,decorations.pathreplacing,positioning,shapes}

\makeatletter
\newcommand\subparagraph{%
  \@startsection{subparagraph}{5}
  {\parindent}
  {3.25ex \@plus 1ex \@minus .2ex}
  {-1em}
  {\normalfont\normalsize\bfseries}}
\makeatother

\usepackage{titlesec}
\let\subparagraph\relax
\titlespacing{\section}{3pt}{4pt plus 2pt minus 1pt}{3pt plus 2pt minus 1pt}
\titlespacing{\subsection}{3pt}{3pt plus 1pt minus 0pt}{2pt plus 1pt minus 0pt}

\usepackage[font=small]{caption}
\usepackage{comment}
\usepackage{epstopdf}
\usepackage{pgfplots}
\usepgfplotslibrary{groupplots}
\pgfplotsset{compat=1.17}

\newcommand{\f}{\mathbf{f}}
\newcommand{\g}{\mathbf{g}}
\newcommand{\h}{\mathbf{h}}

\newcommand{\p}{\mathbf{p}}

\renewcommand{\v}{\mathbf{v}}
\newcommand{\w}{\mathbf{w}}
\newcommand{\x}{\mathbf{x}}
\newcommand{\y}{\mathbf{y}}
\newcommand{\z}{\mathbf{z}}

\newcommand{\0}{\mathbf{0}}

\newcommand{\C}{\mathbf{C}}
\newcommand{\D}{\mathbf{D}}
\newcommand{\E}{\mathbf{E}}

\renewcommand{\H}{\mathbf{H}}
\newcommand{\I}{\mathbf{I}}

\renewcommand{\P}{\mathbf{P}}

\newcommand{\X}{\mathbf{X}}
\newcommand{\Y}{\mathbf{Y}}
\newcommand{\Z}{\mathbf{Z}}

\newcommand{\xib}{\boldsymbol{\xi}}

\newcommand{\Omegab}{\mathbf{\Omega}}

\newcommand{\setB}{\mathcal{B}}
\newcommand{\setC}{\mathcal{C}}

\newcommand{\setG}{\mathcal{G}}

\newcommand{\setK}{\mathcal{K}}
\newcommand{\setL}{\mathcal{L}}

\newcommand{\setN}{\mathcal{N}}

\newcommand{\Real}{\mbox{$\mathbb{R}$}}
\newcommand{\Compl}{\mbox{$\mathbb{C}$}}
\newcommand{\rmF}{\mathrm{F}}

\newcommand{\argmin}{\operatornamewithlimits{argmin}}
\newcommand{\blkdiag}{\mathrm{blkdiag}}

\newcommand{\Diag}{\mathrm{Diag}}

\newcommand{\Exp}{\mathbb{E}}
\newcommand{\herm}{\mathrm{H}}

\renewcommand{\Re}{\mathrm{Re}}

\newcommand{\tran}{\mathrm{T}}

\newtheorem{theorem}{Theorem}

\newtheorem{proposition}{Proposition}
\newtheorem{remark}{Remark}

\newcommand{\bs}{\textnormal{\tiny{BS}}}
\newcommand{\dl}{\textnormal{\tiny{DL}}}

\newcommand{\mse}{\mathrm{MSE}}

\newcommand{\ue}{\textnormal{\tiny{UE}}}
\newcommand{\ul}{\textnormal{\tiny{UL}}}
\newcommand{\ulA}{\textnormal{\tiny{UL-1}}}
\newcommand{\ulB}{\textnormal{\tiny{UL-2}}}
\newcommand{\ulC}{\textnormal{\tiny{UL-3}}}
\newcommand{\sgmse}{\textnormal{\tiny{SG-MSE}}}
\newcommand{\smse}{\textnormal{\tiny{S-MSE}}}

\title{Pilot-Aided Distributed Multi-Group Multicast Precoding Design for Cell-Free Massive MIMO}
\author{Bikshapathi Gouda,~\IEEEmembership{Student Member,~IEEE}, Italo Atzeni,~\IEEEmembership{Member,~IEEE}, \\ and Antti Tölli,~\IEEEmembership{Senior Member,~IEEE}
\thanks{The authors are with the Centre for Wireless Communications, University of Oulu, Finland (e-mail: \{bikshapathi.gouda, italo.atzeni, antti.tolli\}@oulu.fi). This work is supported by the Research Council of Finland (318927 6G~Flagship, 336449 Profi6, and 348396 HIGH-6G) and by the European Commission (101095759 Hexa-X-II). Part of this work was presented at IEEE GLOBECOM 2022~\cite{Gou22}.}}

\begin{document}

\maketitle

\begin{abstract}
We propose fully distributed multi-group multicast precoding designs for cell-free massive multiple-input multiple-output (MIMO) systems with modest training overhead. We target the minimization of the sum of the maximum mean squared errors (MSEs) over the multicast groups, which is then approximated with a weighted sum MSE minimization to simplify the computation and signaling. To design the joint network-wide multi-group multicast precoders at the base stations (BSs) and the combiners at the user equipments (UEs) in a fully distributed fashion, we adopt an iterative bi-directional training scheme with UE- and/or group-specific precoded uplink pilots and group-specific precoded downlink pilots. To this end, we introduce a new group-specific over-the-air uplink training resource that entirely eliminates the need for backhaul signaling for the channel state information (CSI) exchange. The precoders are optimized locally at each BS by means of either best-response or gradient-based updates, and the convergence of the two approaches is analyzed with respect to the centralized implementation with perfect CSI. Finally, numerical results show that the proposed distributed methods greatly outperform conventional cell-free massive MIMO precoding designs that rely solely on local CSI.

\end{abstract}

\section{Introduction}

Emerging shared wireless applications, such as video streaming, vehicular communications, augmented/mixed reality, and wireless coded caching, considerably increase the demand for multicasting services~\cite{Raj20}. The multicast precoding framework was initially developed to transmit a single data stream to a group of user equipments (UEs)~\cite{Sid06}. This was subsequently extended in~\cite{Kar08} to serve several multicast groups with parallel data streams, each transmitted using a group-specific precoder under a rate constraint imposed by the worst UE in the multicast group. The conventional objective considered for the multi-group multicast precoding design is the max-min fairness, according to which the minimum signal-to-interference-plus-noise ratio (SINR) in each multicast group is maximized under a transmit power constraint~\cite{Sid06,Kar08}. For this objective,~\cite{Don20,Mah21} proposed low-complexity methods to design the optimal multi-group multicast precoders. Such precoders have a similar structure to the weighted minimum mean squared error (MMSE) precoder, where the matched filtering (MF) front-end is given by a weighted sum of the effective channels in the multicast group~\cite{Don20}.

The aforementioned works assume perfect channel state information (CSI) at the transmitter. However, in practice, the UE-specific channels need to be estimated. In time division duplexing (TDD) systems with channel reciprocity, this can be done via reverse link measurements, which usually require as many orthogonal pilots as the number of UEs to avoid pilot contamination. The number of orthogonal pilots can be substantially reduced by assigning a common pilot to all the UEs in a multicast group~\cite{Hon13}. Hence, considering the resulting training overhead, using group-specific rather than UE-specific pilots for the multi-group multicast precoding design has the potential to increase the effective rate. The effective performance of multi-group multicasting in massive multiple-input multiple-output (MIMO) systems was analyzed in~\cite{Sad18} under different precoding and pilot assignment strategies. This study was extended in~\cite{Sad18a} to include coexisting unicast and multi-group multicast transmissions. The multi-group multicast precoding design in a coordinated multi-cell scenario was considered in~\cite{Xia13,Son16,Ter17}, where the CSI is assumed to be exchanged among the BSs via backhaul signaling.

Cell-free massive MIMO is an extension of joint transmission coordinated multi-point to a UE-centric approach, where all the BSs jointly serve all the UEs to eliminate the inter-cell interference~\cite{Ngo17,Bjo20a,Raj20}. To facilitate the UE-centric joint processing, the BSs are connected to a central processing unit (CPU) via backhaul links to exchange the UE-specific data and CSI. Most works on cell-free massive MIMO consider simple local precoding strategies, such as MF, local (regularized) zero forcing, and local MMSE precoding~\cite{Bjo20a,Int20,Bjo20}, to circumvent the prohibitive complexity and backhaul signaling of large-scale centralized precoding designs. However, allowing (limited) coordination among the BSs to enable more advanced precoding strategies can provide significant performance gains~\cite{Bjo20,Du21,Mir22,Kal18}. In our previous work~\cite{Atz21,Gou20,Atz20}, we considered a cell-free massive MIMO unicasting scenario and proposed a fully distributed method based on iterative bi-directional training~\cite{Tol19} to design the joint network-wide MMSE precoders locally at each BS. This scheme eliminates the need for backhaul signaling for the CSI exchange altogether and yields a performance close to that of the centralized implementation with perfect CSI.

Cell-free massive MIMO is especially suited for multicasting applications as it improves the rate of the cell-edge UEs and thus reduces the impact of the worst UE in each multicast group. Multi-group multicasting in cell-free massive MIMO systems has been considered, for example, in~\cite{Doa17,Zha19b,Far21}, where MF precoding is used for the data transmission. Equal power allocation among the multicast precoders at each BS was assumed in~\cite{Doa17} to eliminate the need for backhaul signaling for the CSI exchange, whereas the optimal power allocation among the multicast groups was carried out in~\cite{Zha19b,Far21} while assuming limited backhaul signaling.

\subsection{Contribution}

Most works on cell-free massive MIMO multi-group multicasting assume MF precoding to avoid the complexity and backhaul signaling issues associated with the centralized precoding design~\cite{Doa17,Zha19b,Far21}. In this paper, we propose a distributed framework to design the multi-group multicast precoders with low complexity and without any backhaul signaling for the CSI exchange.

We begin by targeting the minimization of the sum of the maximum mean squared errors (MSEs) over the multicast groups, which is referred to in the following as the \textit{sum-group MSE}. This approach achieves absolute MSE fairness within each multicast group, which is dictated by slowly varying dual variables that would need to be exchanged among the BSs via backhaul signaling in the distributed precoding designs. To avoid the resulting backhaul signaling overhead, we approximate the sum-group MSE minimization with a weighted sum MSE minimization, which greatly simplifies the distributed precoding design while only slightly relaxing the MSE fairness requirement. In this regard, we show that the in-built MSE fairness of the weighted sum MSE metric provides a good approximation for the original sum-group MSE metric, especially at high signal-to-noise ratio (SNR). Based on the reformulated problem, we propose a novel framework to design the joint network-wide multi-group multicast precoders at the BSs and the combiners at the UEs in a fully distributed fashion. To this end, we adopt an iterative bi-directional training mechanism~\cite{Tol19} with UE- and/or group-specific precoded uplink pilots and group-specific precoded downlink pilots. The iterative optimization of the precoders is carried out via either best-response or gradient-based updates, and the convergence of the two approaches is analyzed with respect to the centralized implementation with perfect CSI. In our previous work on distributed precoding design for cell-free massive MIMO unicasting~\cite{Atz21}, we introduced a UE-specific over-the-air (OTA) uplink training resource to facilitate the distributed precoding design. In this paper, we propose a new group-specific OTA uplink training resource tailored for the multi-group multicasting scenario, which entirely eliminates the need for backhaul signaling for the CSI exchange and enables the proposed distributed precoding designs with modest training overhead. Moreover, the proposed framework can straightforwardly handle the coexistence of multicasting and unicasting by simply considering individual UEs as separate multicast groups. Numerical results show that the proposed distributed methods bring substantial gains over conventional cell-free massive MIMO precoding designs that rely solely on local CSI. Among the proposed distributed methods, the ones based on group-specific pilots always yield the best effective performance.

The contributions of this paper are summarized as follows.
\begin{itemize}
   \item[$\bullet$] We formulate the multi-group multicast precoding design problem as a sum-group MSE minimization, which is approximated with a weighted sum MSE minimization to avoid the resulting backhaul signaling overhead.
   \item[$\bullet$] We show that the UE-specific rates obtained with the weighted sum MSE minimization asymptotically approximate the ones resulting from the sum-group MSE minimization.
   \item[$\bullet$] We introduce a new group-specific OTA uplink training resource that enables distributed precoding designs.
   \item[$\bullet$] We propose two distributed methods based on iterative bi-directional training with best-response and gradient-based updates, leveraging UE- and/or group-specific pilots, which greatly outperform the reference precoding schemes.
   \item [$\bullet$] We establish that, with perfect CSI, the distributed precoding design with gradient-based updates converges to the same solution as its centralized implementation.
\end{itemize}

Part of this work is included in our conference paper~\cite{Gou22}, which presents the distributed multi-group multicast precoding design with best-response updates.

\subsection{Outline}

The rest of the paper is structured as follows. Section~\ref{sec:SM} introduces the system model for cell-free massive MIMO multi-group multicasting along with the iterative bi-directional training and channel estimation. Section~\ref{sec:problem} describes the sum-group MSE minimization and the approximation with a weighted sum MSE minimization with reference to the centralized implementation with perfect CSI. The proposed distributed multi-group multicast precoding designs with best-response and gradient-based updates are presented in Sections~\ref{sec:distr-perf} and~\ref{sec:distr-imperf} for perfect and imperfect CSI, respectively. Finally, Sections~\ref{sec:num} and~\ref{sec:CONC} provide the numerical results and the concluding remarks, respectively

\subsection{Notation}

Lowercase and uppercase boldface letters denote vectors and matrices, respectively. $(\cdot)^{\tran}$ and $(\cdot)^{\herm}$ are the transpose and Hermitian transpose operators, respectively. $\| \cdot \|$ and $\| \cdot \|_{\rmF}$ represent the Euclidean norm for vectors and the Frobenius norm for matrices, respectively. $\Re[\cdot]$ and $\Exp[\cdot]$ are the real part and expectation operators, respectively. $\I_{L}$ denotes the $L$-dimensional identity matrix and $\0$ represents a zero vector with proper dimension. $\Diag(\cdot)$ and $\blkdiag(\cdot)$ represent diagonal and block-diagonal matrices, respectively. $[a_{1}, \ldots, a_{L}]$ denotes horizontal concatenation, whereas $\{a_{1}, \ldots, a_{L}\}$ and $\{ a_{\ell} \}_{\ell \in \setL}$ represent sets; the latter notation is occasionally relaxed as $\{ a_{\ell} \}$ for brevity. $\setC \setN(0, \sigma^{2})$ is the complex normal distribution with zero mean and variance $\sigma^{2}$. Lastly, $\nabla_{\x}(\cdot)$ denotes the gradient with respect to $\x$, whereas $\mathcal{L}_{(\textrm{P})}(\cdot)$ represents the Lagrangian of optimization problem $(\textrm{P})$.

\section{System Model} \label{sec:SM}

Consider a cell-free massive MIMO system where a set of BSs $\setB \triangleq \{1, \ldots, B\}$ serves a set of UEs $\setK \triangleq \{1, \ldots, K\}$ in the downlink. Each BS and UE are equipped with $M$ and $N$ antennas, respectively. The UEs are divided into a set of non-overlapping multicast groups $\setG \triangleq \{1, \ldots, G\}$, with $\setK_g$ denoting the set of UEs in group~$g \in \setG$.\footnote{The proposed framework and precoding designs are independent of the UE grouping strategies. We assume that the multicast groups are defined by the application layer based on the service requests by the UEs.} In the following, we use $g_{k}$ as the index of the multicast group that contains UE~$k$. The BSs transmit a single data stream to each multicast group, i.e., all the UEs $k \in \setK_g$ are intended to receive the same data symbol~$d_{g}$. Let $\H_{b,k} \in \Compl^{M \times N}$ be the uplink channel matrix between UE~$k \in \setK$ and BS~$b \in \setB$, and let $\w_{b,g} \in \Compl^{M \times 1}$ be the BS-specific precoder used by BS~$b$ for group~$g$. We use $\H_{k} \triangleq [\H_{1,k}^{\tran}, \ldots, \H_{B,k}^{\tran}]^{\tran} \in \Compl^{B M \times N}$ and $\w_{g} \triangleq [\w_{1,g}^{\tran}, \ldots, \w_{B,g}^{\tran}]^{\tran} \in \Compl^{B M \times 1}$ to denote the aggregated uplink channel matrix of UE~$k$ and the aggregated precoder used for group~$g$, respectively, which imply $\H_{k}^{\herm} \w_{g} = \sum_{b \in \setB} \H_{b,k}^{\herm} \w_{b,g}$. We assume the per-BS transmit power constraints $\sum_{g \in \setG} \|\w_{b,g}\|^{2} \leq \rho_{\bs},~\forall b \in \setB$, where $\rho_{\bs}$ denotes the maximum transmit power at each BS. Hence, the signal received at UE~$k$ is given by
\begin{align} \label{eq:y_k}
\y_{k} \triangleq \sum_{b \in \setB} \H_{b,k}^{\herm} \w_{b,g_{k}} d_{g_{k}} \! + \! \sum_{b \in \setB} \sum_{\bar{g} \neq g_{k}} \H_{b,k}^{\herm} \w_{b,\bar{g}} d_{\bar{g}} \! + \! \z_{k} \in \Compl^{N \times 1} \! ,
\end{align}
where $d_{g_k}$ represents the data symbol intended for the group that contains UE~$k$ and $\z_{k} \in \Compl^{N \times 1}$ is the additive white Gaussian noise (AWGN) with i.i.d. $\setC \setN (0, \sigma_{\ue}^{2})$ elements. Upon receiving $\y_{k}$, UE~$k$ obtains a soft estimate of $d_{g}$ by applying the combiner $\v_{k} \in \Compl^{N \times 1}$ and the resulting SINR can be expressed as
\begin{align} \label{eq:SINR_k}
\gamma_{k} \triangleq \frac{|\sum_{b \in \setB} \v_{k}^{\herm} \H_{b,k}^{\herm} \w_{b,g_{k}}|^{2}}{\sum_{\bar g \ne g_{k}} |\sum_{b \in \setB} \v_{k}^{\herm} \H_{b,k}^{\herm} \w_{b,\bar{g}}|^{2} + \sigma_{\ue}^{2} \| \v_{k} \|^{2}}.
\end{align}
Finally, the sum of the rates over the multicast groups, which is referred to in the following as the \textit{sum-group rate}, is given by $R \triangleq \sum_{g \in \setG} R_g$, where $R_g$ is the rate of group~$g$ defined as
\begin{align} \label{eq:R}
R_g \triangleq \min_{k \in \setK_g} \log_{2}(1 + \gamma_{k}) \quad \textrm{[bps/Hz]}.
\end{align}
Note that \eqref{eq:R}, which is based on the SINR expression in \eqref{eq:SINR_k}, represents an upper bound on the system performance that assumes perfectly estimated SINRs for given precoders and combiners.\footnote{This can be computed at the BSs directly if perfect global CSI is available~\cite{Cai18} (at least the effective uplink channels $\{ \H_{b,k} \v_{k} \}_{b \in \setB, k \in \setK}$) or iteratively by allowing the UEs to send channel quality indicator feedback of their estimated SINRs.} In Section~\ref{sec:num}, we use this metric to evaluate the proposed distributed multi-group multicast precoding designs.

In this paper, we aim to design the joint network-wide multi-group multicast precoders at the BSs and the combiners at the UEs in a fully distributed fashion assuming an ideal TDD setting with channel reciprocity between uplink and downlink. To this end, we adopt an iterative bi-directional training scheme that relies on estimating the effective uplink and downlink channels via precoded pilots, as discussed in detail in the following section.

\subsection{Pilot-Aided Channel Estimation and Iterative Bi-Directional Training} \label{sec:SM_est}
  
The centralized precoding design (considered as reference scheme and described in Section~\ref{sec:problem_ref}) involves the transmission of antenna-specific uplink pilots, by which each BS estimates the antenna-specific uplink channels.

\smallskip

\textit{\textbf{Antenna-specific uplink channel estimation (UL).}} The estimation of the uplink channel $\H_{b,k}$ involves $N$ antenna-specific uplink pilots for UE~$k$. In this context, let $\P_{k}^{\ul} \in \Compl^{\tau^{\ul} \times N}$ be the uplink pilot matrix of UE~$k$, with $\|\P_{k}^{\ul}\|_{\rmF}^{2} = \tau^{\ul} N,~\forall k \in \setK$. Moreover, let $\rho_{\ue}$ denote the maximum transmit power at each UE. Each UE~$k$ synchronously transmits its pilot matrix $\P_{k}^{\ul}$,~i.e.,
\begin{align} \label{eq:X_k_ul}
\X_{k}^{\ul} \triangleq \sqrt{\beta^{\ul}} (\P_{k}^{\ul})^{\herm} \in \Compl^{N \times \tau^{\ul}},
\end{align}
where the power scaling factor $\beta^{\ul} \triangleq \frac{\rho_{\ue}}{N}$ (equal for all the UEs) ensures that $\X_{k}^{\ul}$ complies with the per-UE transmit power constraint. Then, the signal received at BS~$b$ is given by
\begin{align} \label{eq:Y_b_ul}
\Y_{b}^{\ul} & \triangleq \sum_{k \in \setK} \H_{b,k} \X_{k}^{\ul} + \Z_{b}^{\ul} \\
& = \sqrt{\beta^{\ul}} \sum_{k \in \setK} \H_{b,k} (\P_{k}^{\ul})^{\herm} + \Z_{b}^{\ul} \in \Compl^{M \times \tau^{\ul}},
\end{align}
where $\Z_{b}^{\ul}$ is the AWGN with i.i.d. $\setC \setN (0, \sigma_{\bs}^{2})$ elements. Finally, the least-squares (LS) estimate of $\H_{b,k}$ is
\begin{align} \label{eq:H_bk_hat_orth}
\hat{\H}_{b,k} & \triangleq \frac{1}{\tau^{\ul} \sqrt{\beta^{\ul}}} \Y_{b}^{\ul} \P_{k}^{\ul} \\
& = \H_{b,k} + \frac{1}{\tau^{\ul}} \sum_{\bar{k}  \ne k} \H_{b,\bar{k}} (\P_{\bar{k}}^{\ul})^{\herm} \P_{k}^{\ul} + \frac{1}{\tau^{\ul} \sqrt{\beta^{\ul}}} \Z_{b}^{\ul} \P_{k}^{\ul},
\end{align}
where the last equality holds if $(\P_{k}^{\ul})^{\herm} \P_{k}^{\ul} = \tau^{\ul} \I_{N}$, i.e., if there is no pilot contamination among the antennas of UE~$k$.

\smallskip

\begin{figure}[t!]
\begin{center}
\begin{tikzpicture}

\tikzstyle{rect_round}=[draw, rectangle, rounded corners, text centered, minimum height=3mm];

\footnotesize

\draw node[align=center] (xxx) {};
\draw node[rect_round, left of=xxx, node distance=25mm, align=center, anchor=center] (UE) {\normalsize UE~$k$};
\draw node[rect_round, right of=xxx, node distance=25mm, align=center, anchor=center] (BS) {\normalsize BS~$b$};

\draw node[below of=UE, node distance=7.5mm, align=center, anchor=center] (UE_0) {initialize $\v_{k}$};
\draw node[below of=UE, node distance=10mm, align=center, anchor=center] (UE_0_south) {};
\draw node[below of=BS, node distance=15mm, align=center, anchor=center] (BS_1_north) {};
\draw node[below of=BS, node distance=17.5mm, align=center, anchor=center] (BS_1) {update $\w_{b}$};
\draw node[below of=BS, node distance=20mm, align=center, anchor=center] (BS_1_south) {};
\draw node[below of=UE, node distance=25mm, align=center, anchor=center] (UE_1_north) {};
\draw node[below of=UE, node distance=27.5mm, align=center, anchor=center] (UE_1) {update $\v_{k}$};
\draw node[below of=UE, node distance=30mm, align=center, anchor=center] (UE_1_south) {};
\draw node[below of=BS, node distance=35mm, align=center, anchor=center] (BS_2_north) {};
\draw node[below of=BS, node distance=37.5mm, align=center, anchor=center] (BS_2) {update $\w_{b}$};
\draw node[below of=BS, node distance=40mm, align=center, anchor=center] (BS_2_south) {};
\draw node[below of=UE, node distance=45mm, align=center, anchor=center] (UE_2_north) {};

\begin{scope}[>=latex]
\draw[->] ($(UE_0_south)$) -- ($(BS_1_north)$) node[midway, above, sloped] {precoded uplink pilots};
\draw[->] ($(BS_1_south)$) -- ($(UE_1_north)$) node[midway, above, sloped] {precoded downlink pilots};
\draw[->] ($(UE_1_south)$) -- ($(BS_2_north)$) node[midway, above, sloped] {precoded uplink pilots};
\draw[->] ($(BS_2_south)$) -- ($(UE_2_north)$) node[midway, above, sloped] {data symbols};
\end{scope}

\draw[decorate,decoration={brace,amplitude=10pt}] ($(BS_1_north)+(7.5mm,0mm)$) -- ($(BS_2_north)+(7.5mm,0mm)$) node[align=left, midway, xshift=12.5mm] {\footnotesize repeat until \\ a predefined \\ termination \\ criterion is \\ satisfied};

\end{tikzpicture}
\end{center}
\caption{Schematic representation of iterative bi-directional training in a single-UE, single-BS setting. To transmit the precoded uplink pilots, UE~$k$ uses $\v_{k}$ as precoder.}
\label{fig:BDT}
\end{figure}
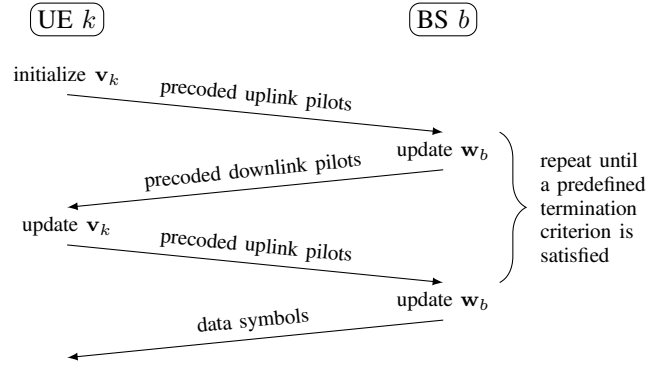

On the other hand, the proposed distributed precoding designs and the local precoding designs (also considered as reference schemes and described in Appendix~\ref{app:A3}) are based on \textit{iterative bi-directional training}, whereby the precoders at the BSs and the combiners at the UEs are updated iteratively by means of uplink and downlink pilot-aided channel estimation~\cite{Jay18,Kal18,Tol19}. Specifically, each bi-directional training iteration involves:
\begin{itemize}
\item[\textit{i)}] The transmission of UE- and/or group-specific precoded uplink pilots from all the UEs, by which each BS estimates the UE- and/or group-specific effective uplink channels and updates its precoders;
\item[\textit{ii)}] The transmission of precoded downlink pilots from all the BSs, by which each UE estimates its effective downlink channel and updates its combiner.
\end{itemize}
Iterative bi-directional training can reduce the training overhead compared with antenna-specific uplink channel estimation for multi-antenna UEs. More importantly, it eliminates the need for centralized precoding design since each BS (resp. UE) can update its precoder (resp. combiner) based on the effective uplink (resp. downlink) channel estimation. A schematic representation of iterative bi-directional training in a single-UE, single-BS setting is provided in Figure~\ref{fig:BDT}. In the following, we describe the different \textit{existing} types of pilot-aided channel estimation that are adopted within the iterative bi-directional training, which will be heavily utilized in Sections~\ref{sec:problem_ref} and~\ref{sec:distr-imperf} as well as in Appendix~\ref{app:A3}. In Section~\ref{sec:distr-imperf}, we further introduce a \textit{new} group-specific OTA uplink training resource tailored for the multi-group multicasting scenario, which entirely eliminates the need for backhaul signaling for the CSI exchange and enables the proposed distributed precoding designs with modest training overhead.

\smallskip

\textit{\textbf{UE-specific effective uplink channel estimation (UL-1).}} Let $\h_{b,k} \triangleq \H_{b,k} \v_{k} \in \Compl^{M \times 1}$ be the effective uplink channel between UE~$k$ and BS~$b$, and let $\p_{k}^{\ulA} \in \Compl^{\tau^{\ulA} \times 1}$ denote the uplink pilot of UE~$k$, with $\|\p_{k}^{\ulA}\|^{2} = \tau^{\ulA},~\forall k \in \setK$. Each UE~$k$ synchronously transmits its pilot $\p_{k}^{\ulA}$ using its scaled combiner $\v_{k}$ as precoder, i.e.,
\begin{align} \label{eq:X_k_ul1}
\X_{k}^{\ulA} \triangleq \sqrt{\beta^{\ulA}} \v_{k} (\p_{k}^{\ulA})^{\herm} \in \Compl^{N \times \tau^{\ulA}},
\end{align}
where the power scaling factor $\beta^{\ulA}$ (equal for all the UEs) ensures that $\X_{k}^{\ulA}$ complies with the per-UE transmit power constraint. Then, the signal received at BS~$b$ is given by
\begin{align} \label{eq:Y_b_ul1}
\Y_{b}^{\ulA} & \triangleq \sum_{k \in \setK} \H_{b,k} \X_{k}^{\ulA} + \Z_{b}^{\ulA} \\
& = \sqrt{\beta^{\ulA}} \sum_{k \in \setK} \h_{b,k} (\p_{k}^{\ulA})^{\herm} + \Z_{b}^{\ulA} \in \Compl^{M \times \tau^{\ulA}},
\end{align}
where $\Z_{b}^{\ulA}$ is the AWGN with i.i.d. $\setC \setN (0, \sigma_{\bs}^{2})$ elements. Finally, the LS estimate of $\h_{b,k}$ is
\begin{align} \label{eq:h_bk_hat}
\hat{\h}_{b,k} & \triangleq \frac{1}{\tau^{\ulA} \sqrt{\beta^{\ulA}}} \Y_{b}^{\ulA} \p_{k}^{\ulA} \\
& = \h_{b,k} \! + \! \frac{1}{\tau^{\ulA}} \sum_{\bar{k} \ne k} \h_{b,\bar{k}} (\p_{\bar{k}}^{\ulA})^{\herm} \p_{k}^{\ulA} \! + \! \frac{1}{\tau^{\ulA} \sqrt{\beta^{\ulA}}} \Z_{b}^{\ulA} \p_{k}^{\ulA} \! .
\end{align}

\smallskip

\textit{\textbf{Group-specific effective uplink channel estimation (UL-2).}} In the antenna-specific and UE-specific channel estimations described above, the BSs may apply UE-specific weights to the channel estimates to promote fairness among the UEs in a multicast group. On the contrary, in the group-specific channel estimation, any UE-specific weights must be already incorporated during the pilot transmission. Accordingly, let $\omega_k$ be the weight of UE~$k$ and let $\f_{b,g} \triangleq \sum_{ k \in \setK_g} \omega_k \H_{b,k} \v_{k} \in \Compl^{M \times 1}$ denote the effective uplink channel between $\setK_{g}$ and BS~$b$. Furthermore, let $\p_{g}^{\ulB} \in \Compl^{\tau^{\ulB} \times 1}$ be the uplink pilot of group~$g$, with $\|\p_{g}^{\ulB}\|^{2} = \tau^{\ulB},~\forall g \in \setG$. Each UE~$k$ synchronously transmits its pilot $\p_{g_{k}}^{\ulB}$ using its scaled combiner $\v_{k}$ as precoder, i.e.,
\begin{align} \label{eq:X_g_ul2}
\X_{k}^{\ulB} \triangleq \sqrt{\beta^{\ulB}} \omega_k \v_{k} (\p_{g_{k}}^{\ulB})^{\herm} \in \Compl^{N \times \tau^{\ulB}},
\end{align}
where the power scaling factor $\beta^{\ulB}$ (equal for all the UEs) ensures that $\X_{k}^{\ulB}$ complies with the per-UE transmit power constraint. Then, the signal received at BS~$b$ is given by
\begin{align} \label{eq:Y_b_ul2}
\Y_{b}^{\ulB} & \triangleq \sum_{k \in \setK} \H_{b,k} \X_{k}^{\ulB} + \Z_{b}^{\ulB} \\
& = \sqrt{\beta^{\ulB}} \sum_{g \in \setG} \f_{b,g} (\p_{g}^{\ulB})^{\herm} + \Z_{b}^{\ulB} \in \Compl^{M \times \tau^{\ulB}},
\end{align}
where $\Z_{b}^{\ulB}$ is the AWGN with i.i.d. $\setC \setN (0, \sigma_{\bs}^{2})$ elements. Finally, the LS estimate of $\f_{b,g}$ is
\begin{align} \label{eq:f_bk_hat}
\hat{\f}_{b,g} & \triangleq \frac{1}{\tau^{\ulB} \sqrt{\beta^{\ulB}}} \Y_{b}^{\ulB} \p_{g}^{\ulB} \\
& = \f_{b,g} + \frac{1}{\tau^{\ulB}} \sum_{\bar{g} \ne g} \f_{b,\bar{g}} (\p_{\bar{g}}^{\ulB})^{\herm} \p_{g} + \frac{1}{\tau^{\ulB} \sqrt{\beta^{\ulB}}} \Z_{b}^{\ulB} \p_{g}^{\ulB}.
\end{align}

\smallskip

\textit{\textbf{Effective downlink channel estimation (DL).}} Let $\g_{k} \triangleq \sum_{b \in \setB} \H_{b,k}^{\herm} \w_{b,g} \in \Compl^{N \times 1}$ be the effective downlink channel between all the BSs and UE~$k$. Moreover, let $\p_{g}^{\dl} \in \Compl^{\tau^{\dl} \times 1}$ denote the downlink pilot of group~$g$, with $\|\p_{g}^{\dl}\|^{2} = \tau^{\dl},~\forall g \in \setG$. Each BS~$b$ synchronously transmits a superposition of the pilots $\{\p_{g}^{\dl}\}_{g \in \setG}$ after precoding them with the corresponding precoders $\{\w_{b,g}\}_{g \in \setG}$, i.e.,
\begin{align} \label{eq:X_b_dl}
\X_{b}^{\dl} \triangleq \sum_{g \in \setG} \w_{b,g} (\p_{g}^{\dl})^{\herm} \in \Compl^{M \times \tau^{\dl}}.
\end{align}
Then, the signal received at UE~$k$ is given by
\begin{align} \label{eq:Y_k_dl}
\Y_{k}^{\dl} & \triangleq \sum_{b \in \setB} \H_{b,k}^{\herm} \X_{b}^{\dl} + \Z_{k}^{\dl} \\
&= \sum_{b \in \setB} \sum_{g \in \setG} \H_{b,k}^{\herm} \w_{b,g} (\p_{g}^{\dl})^{\herm} + \Z_{k}^{\dl} \in \Compl^{N \times \tau^{\dl}},
\end{align}
where $\Z_{k}^{\dl}$ is the AWGN with i.i.d. $\setC \setN (0, \sigma_{\ue}^{2})$ elements. Finally, the LS estimate of $\g_{k}$ is
\begin{align} \label{eq:g_k_hat}
\hat{\g}_{k} & \triangleq \frac{1}{\tau^{\dl}} \Y_{k}^{\dl} \p_{g_{k}}^{\dl} \\
& = \g_{k} + \frac{1}{\tau^{\dl}} \sum_{b \in \setB} \sum_{\bar{g} \ne g_{k}} \H_{b,k}^{\herm} \w_{b,\bar{g}} (\p_{\bar{g}}^{\dl})^{\herm} \p_{g_{k}}^{\dl} + \frac{1}{\tau^{\dl}} \Z_{k}^{\dl} \p_{g_{k}}^{\dl}.
\end{align}

\smallskip

Note that all the above pilot-aided channel estimation schemes can be implemented with arbitrary pilots and, hence, any possible pilot contamination is implicitly accounted for.

\section{Problem Formulation} \label{sec:problem}

The goal of this paper is to propose fully distributed multi-group multicast precoding designs for cell-free massive MIMO systems based on the MMSE criterion. In this section, we establish the basis for the distributed precoding design by considering the centralized implementation with perfect CSI. First, in Section~\ref{sec:problem_sum-groupMSE}, we focus on the sum-group MSE minimization and identify several practical challenges with its distributed implementation. Then, in Section~\ref{sec:problem_sumMSE}, we approximate the sum-group MSE minimization with a weighted sum MSE minimization, based on which we develop the proposed distributed precoding designs presented in Sections~\ref{sec:distr-perf} and~\ref{sec:distr-imperf} with perfect and imperfect CSI, respectively.

\subsection{Sum-Group MSE Minimization} \label{sec:problem_sum-groupMSE}

The sum-group MSE minimization achieves absolute MSE fairness within each multicast group through the min-max MSE criterion subject to the per-BS transmit power constraints. Accordingly, the precoders and combiners are optimized by solving
\begin{align} \label{eq:probForHi}
\begin{array}{cl}
\displaystyle \underset{\{\w_g, \v_{k}\}}{\mathrm{minimize}} & \displaystyle \sum_{g \in \setG} \max_{k \in \setK_g} \mse_k \\
\mathrm{s.t.} & \displaystyle \sum_{g \in \setG} \| \E_{b} \w_{g}\|^{2} \leq \rho_{\bs}, \quad \forall b \in \setB,
\end{array}
\end{align}
where $\mse_k$ is the MSE of UE~$k$ defined as
\begin{align} \label{eq:MSE_k}
\mse_{k} & \triangleq \Exp \big[ |\v_{k}^{\herm} \y_{k} - d_{g_k}|^{2} \big] \\
& = \sum_{g\in \setG} | \v_{k}^{\herm} \H_{k}^{\herm} \w_{{g}} |^{2} \! - \! 2 \Re [ \v_{k}^{\herm} \H_{k}^{\herm} \w_{g_k}] + \sigma_{\ue}^{2} \| \v_{k} \|^{2} \! + \! 1
\end{align}
and $\E_{b} \in \Real^{M \times B M}$ is a selection matrix such that $\E_b \w_g = \w_{b,g}$. The problem in \eqref{eq:probForHi}
is convex with respect to either the precoders or the combiners but not jointly convex with respect to both. Hence, we use \textit{alternating optimization}, whereby the precoders are optimized for fixed combiners and vice versa in an iterative best-response fashion. Before describing each step of the alternating optimization, let us define $t_g \triangleq \max_{k \in \setK_g} \mse_k$ and rewrite \eqref{eq:probForHi} in epigraph form as
\begin{align} \label{eq:probFor}
\begin{array}{cl}
\displaystyle \underset{\{\w_g, \v_{k}, t_g\}}{\mathrm{minimize}} & \displaystyle \sum_{g \in \setG} t_g \\
\mathrm{s.t.} & \displaystyle \mu_{k} : \mse_{k} \le t_{g}, \quad \forall k \in \setK_g,~\forall g \in \setG \\
& \displaystyle \lambda_{b} : \sum_{g \in \setG} \| \E_{b} \w_{g}\|^{2} \leq \rho_{\bs}, \quad \forall b \in \setB.
\end{array}
\end{align}

\smallskip

\textit{\textbf{Optimization of the combiners.}} For a fixed set of precoders $\{\w_{g}\}_{g \in \setG}$, the combiners $\{\v_{k}\}_{k \in \setK}$ are optimized by solving the following convex problem:
\begin{align} \label{eq:probUE}
\begin{array}{cl}
\displaystyle \underset{\{\v_{k}, t_g\}}{\mathrm{minimize}} & \displaystyle \sum_{g \in \setG} t_g \\
\mathrm{s.t.} & \displaystyle \mse_{k} \le t_{g}, \quad \forall k \in \setK_g,~\forall g \in \setG.
\end{array}
\end{align}
The Lagrangian of \eqref{eq:probUE} can be written as
\begin{align}\label{eq:lag_ue}
\mathcal{L}_{\eqref{eq:probUE}}\big(\{\v_k, t_g, \mu_{k}\}\big) \triangleq  \sum_{g \in \setG} t_g + \sum_{k \in \setK}  \mu_{k} \big(\mse_{k} - t_{g_k}\big),
\end{align}
where $\mu_{k}$ is the dual variable corresponding to each per-UE MSE constraint in \eqref{eq:probFor}. Note that the optimal $\{\mu_{k}\}_{k \in \setK}$ are such that the MSE objectives of the UEs in a multicast group are equal. For example, if UE~$k$ is subject to poor channel conditions, the optimal $\mu_{k}$ will be large to force the reduction of its MSE objective. Then, the optimal $\v_{k}$ is obtained by setting $\nabla_{\v_k} \mathcal{L}_{\eqref{eq:probUE}}\big(\{\v_k, t_g, \mu_{k}\}\big) = \0$, which yields
\begin{align}\label{eq:uebf}
\v_{k} = \bigg( \sum_{ g \in \setG}\H_{k}^{\herm} \w_{g} \w_{ g}^{\herm} \H_{k} + \sigma_{\ue}^{2} \I_{N} \bigg)^{-1}\H_{k}^{\herm} \w_{g_k}.
\end{align}

\smallskip

\textit{\textbf{Optimization of the precoders.}} For a fixed set of combiners $\{\v_{k}\}_{k \in \setK}$, the precoders $\{\w_{g}\}_{g \in \setG}$ are optimized by solving the following convex problem:
\begin{align} \label{eq:probBS}
\begin{array}{cl}
\displaystyle \underset{{\{\w_g, t_g\}}}{\mathrm{minimize}} &  \displaystyle \sum_{g \in \setG} t_g \\
\mathrm{s.t.} & \displaystyle \mse_{k} \le t_{g}, \quad \forall k \in \setK_g,~\forall g \in \setG \\
& \displaystyle \sum_{g \in \setG} \| \E_{b} \w_{g}\|^{2} \leq \rho_{\bs}, \quad \forall b \in \setB,
\end{array}
\end{align}
which can be solved, e.g., via CVX~\cite{cvx14}. Alternatively, one can resort to the Karush–Kuhn–Tucker (KKT) conditions, which also conveniently reveal the optimal multi-group multicast precoding structure. In this regard, the Lagrangian of \eqref{eq:probBS} can be written as
\begin{align}\label{eq:lagr_bs}
\mathcal{L}_{\eqref{eq:probBS}} \big( \{\w_g, t_g, \mu_{k}, \lambda_b\} \big) & \triangleq \sum_{g \in \setG} t_g + \sum_{k \in \setK} \mu_{k} \big(\mse_{k} - t_{g_k}\big) \nonumber \\
& \phantom{=} \ + \sum_{b \in \setB} \lambda_{b} \bigg( \sum_{g \in \setG} \| \E_{b} \w_{g}\|^{2} - \rho_{\bs} \bigg),
\end{align}
where $\lambda_b$ is the dual variable corresponding to each per-BS transmit power constraint in \eqref{eq:probFor}. Then, the optimal $\w_{g}$ is obtained by setting $\nabla_{\w_g} \mathcal{L}_{\eqref{eq:probBS}}\big(\{\w_g, t_g, \mu_{k}, \lambda_b\}\big) = \0$, which yields
\begin{align}\label{eq:bsbf}
\w_{g} = \bigg( \! \sum_{ k \in \setK}\mu_{ k}\H_{ k}\v_{ k} \v_{ k}^{\herm} \H_{k} ^{\herm} \! + \! \sum_{b \in \setB} \lambda_{b} \E_{b}^{\herm} \E_{b} \! \bigg)^{-1} \! \! \sum_{ k \in \setK_g} \mu_{k} \H_{k} \v_{k}.
\end{align}
The above expression of $\w_g$ depends on the dual variables $\{\mu_k\}_{k \in \setK}$ and $\{\lambda_b\}_{b \in \setB}$. Such dual variables can be updated iteratively using the sub-gradient method as detailed in Appendix~\ref{app:A1}~\cite{Mah21,Kom13}, and their values after convergence are finally used in \eqref{eq:bsbf}. 

\smallskip

From the expression of the aggregated precoder $\w_{g}$ in \eqref{eq:bsbf}, it is evident that the BS-specific precoders $\{\w_{b,g}\}_{b \in \setB}$ also rely on the dual variables $\{\mu_k\}_{k \in \setK}$. To compute each $\w_{b,g}$ locally at BS~$b$, extensive backhaul signaling is required to iteratively update the dual variables $\{\mu_{k}\}_{k \in \setK}$ either at the CPU or at each BS in parallel. To simplify the distributed precoding design, we propose to relax the absolute MSE fairness requirement within each multicast group, which leads to a weighted sum MSE minimization. In the following section, we describe the reformulated problem and the corresponding centralized precoding design with perfect CSI.

\subsection{Weighted Sum MSE Minimization} \label{sec:problem_sumMSE}

To circumvent the shortcomings of the original problem formulation described in Section~\ref{sec:problem_sum-groupMSE}, we approximate the sum-group MSE objective in \eqref{eq:probForHi} with a weighted sum MSE objective. Accordingly, the precoders and combiners are optimized by solving
\begin{align} \label{eq:EqvprobForHi}
\begin{array}{cl}
\displaystyle \underset{\{\w_g, \v_{k}\}}{\mathrm{minimize}} & \displaystyle \sum_{k \in \setK} \omega_k\mse_k \\
\mathrm{s.t.} &  \displaystyle \lambda_{b} : \sum_{g \in \setG} \| \E_{b} \w_{g}\|^{2} \leq \rho_{\bs}, \quad \forall b \in \setB, 
\end{array}
\end{align}
where we recall that $\omega_k$ is the weight of UE~$k$. This choice stems from the fact that the weighted sum MSE metric provides some in-built MSE fairness among all the UEs. Since the problem in \eqref{eq:EqvprobForHi} is convex with respect to either the precoders or the combiners but not jointly convex with respect to both, we use alternating optimization as in the previous section. For a fixed set of combiners $\{\v_k\}_{k \in \setK}$, the precoders $\{\w_g\}_{g \in \setG}$ can be optimized, e.g., via CVX~\cite{cvx14} or by resorting to the KKT conditions. In this regard, the Lagrangian of \eqref{eq:EqvprobForHi} can be written as
\begin{align}\label{eq:lagbsmse}
\mathcal{L}_{\eqref{eq:EqvprobForHi}} \big( \{ \w_g,\lambda_b \} \big) & \triangleq \sum_{k \in \setK} \omega_{k} \mse_{k} \nonumber \\
& \phantom{=} \ + \sum_{b \in \setB} \lambda_{b} \bigg( \sum_{g \in \setG} \| \E_{b} \w_{g}\|^{2} - \rho_{\bs} \bigg).
\end{align}
Then, the optimal $\w_{g}$ is obtained by setting $\nabla_{\w_g} \mathcal{L}_{\eqref{eq:EqvprobForHi}}\big(\{ \w_g,\lambda_b\}\big)=\0$, which yields
\begin{align}\label{eq:bsbf_mse}
\w_{g} = \bigg( \! \sum_{ k \in \setK}\omega_{ k}\H_{ k}\v_{ k} \v_{ k}^{\herm} \H_{ k}^{\herm} \! + \! \sum_{b \in \setB} \lambda_{b} \E_{b}^{\herm} \E_{b} \! \bigg)^{-1} \! \!  \sum_{ k \in \setK_g}\omega_{k} \H_{k} \v_{k}.
\end{align}
It is straightforward to notice the resemblance between \eqref{eq:bsbf_mse} and \eqref{eq:bsbf}. If the optimal dual variables $\{\mu_k\}_{k \in \setK}$ of the sum-group MSE minimization were known in advance, one could replace the weights $\{\omega_k\}_{k \in \setK}$ in \eqref{eq:bsbf_mse} with the optimal $\{ \mu_k\}_{k \in \setK}$ at each alternating optimization iteration, which would lead to the same solution of \eqref{eq:bsbf}. However, the optimal $\{\mu_k\}_{k \in \setK}$ cannot be known in advance. Moreover, tuning the weights to match the dual variables at each alternating optimization iteration would generate the same complexity and backhaul signaling overhead of the original sum-group MSE minimization.\footnote{To promote fairness within each multicast group with reduced complexity and backhaul signaling overhead, the values of $\{\omega_k\}_{k \in \setK}$ may be updated less frequently, e.g., based on SINR reporting from the UEs.}

To simplify the distributed precoding design, we consider the sum MSE minimization with fixed UE-specific weights, which can be assigned to promote fairness or priority within each multicast group based on prior information, e.g., about their channel conditions. Without loss of generality, we fix equal weights for all the UEs, i.e., $\omega_k = \omega,~\forall k \in \setK$, a choice justified by the uniform service provisioning of cell-free massive MIMO systems. Hence, in the following, we refer to \eqref{eq:EqvprobForHi} simply as sum MSE minimization. Though slightly suboptimal, as demonstrated later, this approach leads to much simpler computation and signaling, and is characterized by faster convergence. Note that, especially at high SNR, the UE-specific rates derived from the sum MSE minimization are close to those obtained with the sum-group MSE minimization. This is formalized in Proposition~\ref{pro:1}. Furthermore, for a fixed set of precoders $\{\w_g\}_{g \in \setG}$, the optimal combiners $\{\v_k\}_{k \in \setK}$ for \eqref{eq:EqvprobForHi} are again obtained as in \eqref{eq:uebf} based on the effective downlink channel estimation described in Section~\ref{sec:SM_est}.

\begin{proposition} \label{pro:1}
As $\rho_{\bs} \to \infty$, i.e., at high SNR, the UE-specific rates obtained with the sum MSE minimization in \eqref{eq:EqvprobForHi} asymptotically approximate the ones resulting from the sum-group MSE minimization in \eqref{eq:probForHi}.
\end{proposition}

\begin{IEEEproof}
Without loss of generality, let us consider a single BS and let us define $c_{k \bar k} \triangleq \frac{{\v_k}^{\herm} {\H_k}^{\herm} \w_{g_{\bar k}}}{\|\v_k\| \|\w_{g_{\bar k}}\| }$. Assuming that UE~$k$ adopts the MMSE combiner in \eqref{eq:uebf}, its MSE can be expressed as $\mse_k = \frac{1}{1+\gamma_k}$ (cf. \eqref{eq:MSE_k}). As $\rho_{\bs} \to \infty$, the precoder in \eqref{eq:bsbf_mse} approaches a solution similar to zero forcing, i.e., $\w_g$ lies in the nullspace of the effective uplink channels of the UEs $k \notin \setK_g$ and matched towards the superposition of the effective uplink channels of the UEs $k \in \setK_g$. Thus, considering UE~$\bar{k} \notin \setK_{g_{k}}$, the inner product between $\w_{g_{\bar{k}}}$ and the effective uplink channel of UE~$k$ tends to zero, which leads to $c_{k \bar k} \to 0,~\forall \bar{k} \notin \setK_{g_{k}}$. In this context, all the UEs experience high SINR, and the SINR of UE~$k$ can be approximated as~(cf. \eqref{eq:SINR_k})
\begin{align}\label{eq:snr_UE_k}
\gamma_{k} \approx \frac{p_{g_k} c_{kk}^2}{\sigma^2_{\ue}},
\end{align}
where $p_g$ is the transmit power allocated to group~$g$. Finally, when $\omega_k=\omega,~\forall k \in \setK$, the sum MSE minimization in \eqref{eq:EqvprobForHi} reduces to the following power allocation problem:
\begin{align} \label{eq:appxsmse}
\begin{array}{cl}
\displaystyle \underset{{\{p_{g}\}}}{\mathrm{minimize}} & \displaystyle \sum_{k \in \setK} \frac{\sigma^2_{\ue}}{p_{g_k} c_{kk}^2}\\
\mathrm{s.t.} &  \displaystyle \sum_{g \in \setG} p_{g} \leq \rho_{\bs}. 
\end{array}
\end{align}
From the KKT conditions detailed in Appendix~\ref{app:A2}, we obtain the optimal $p_g$ as
\begin{equation}
p_g = \rho_{\bs} \frac{u_g}{\sum_{\bar g \in \setG} u_{\bar g}},
\end{equation}
with $u_{g} \triangleq \sqrt{\sum_{k \in \setK_g} \frac{\sigma_{\ue}^2}{c_{kk}^2}}$. Consequently, the rate difference between UE~$k$ and UE~$\bar{k}$ at high SNR can be written as 
\begin{align} \label{eq:ratediff}
\big| \log_{2}(\gamma_{k}) - \log_{2}(\gamma_{\bar k}) \big|  
= \bigg|\log_{2} \bigg(\frac{u_{g_k} c_{kk}^2}{u_{g_{\bar k}} c_{\bar k \bar k}^2}\bigg) \bigg|,
\end{align}
which is independent of $\rho_{\bs}$. This suggests that all the UE-specific rates increase uniformly with the transmit power. Considering the MSE fairness requirement of \eqref{eq:probFor}, it follows that the rate of UE~$k \in \setK_g$ obtained with the sum-group MSE minimization lies within the minimum and the maximum rates among all the UEs $\bar{k} \in \setK_g$ obtained with the sum MSE minimization, i.e.,
\begin{equation}
\min_{\bar k \in \setK_g} R^{\smse}_{\bar k} \le R^{\sgmse}_k \le \max_{\bar k \in \setK_g} R^{\smse}_{\bar k}, \quad \forall k \in \setK_g,
\end{equation}
where $R^{\smse}_k$ and $R^{\sgmse}_k$ indicate the rates of UE~$k$ obtained with the sum MSE minimization and with the sum-group MSE minimization, respectively. The asymptotic approximation of the normalized difference between $R^{\smse}_k$ and $R^{\sgmse}_k$ is given by
\begin{align}
& \lim_{\rho_{\bs}\to\infty} \frac{\big|R^{\smse}_k - R^{\sgmse}_k \big|} {R^{\smse}_k} \nonumber \\
& \leq \lim_{\rho_{\bs}\to\infty} \frac{\big|\min_{\bar k \in \setK_g} R^{\smse}_{\bar k} - \max_{\bar k \in \setK_g} R^{\smse}_{\bar k} \big|} {{R^{\smse}_k}} \\
& \simeq \lim_{\rho_{\bs}\to\infty} \frac{\big| \log_{2}( \min_{\bar k \in \setK_g} u_{g_{\bar k}} c_{\bar k \bar k}^2) - \log_{2}(\max_{\bar k \in \setK_g} u_{g_{\bar k}} c_{\bar k \bar k}^2)\big|} {\log_{2}(\rho_{\bs} u_{g_k} {{c_{kk}}^2)} - \log_{2}(\sigma^2_{\ue}\sum_{\bar g \in \setG} u_{\bar g})} \\
& \to 0, \ \forall k \in \setK_g.
\end{align}
Hence, at high SNR, the UE-specific rates obtained with the sum MSE minimization asymptotically approximate the ones resulting from the sum-group MSE minimization.
\end{IEEEproof}

\smallskip

In the rest of the paper, we focus on the sum MSE minimization in \eqref{eq:EqvprobForHi} to design the multi-group multicast precoders. The proposed distributed precoding designs presented in Sections~\ref{sec:distr-perf} and~\ref{sec:distr-imperf} with perfect and imperfect CSI, respectively, are compared with different reference schemes, namely: \textit{i)}~the centralized precoding design presented in Section~\ref{sec:problem_ref}, which is referred to in the following as the \textit{Centralized}; and \textit{ii)}~the local precoding designs based on MMSE and MF described in Appendix~\ref{app:A3}, which are referred to in the following as the \textit{Local~MMSE} and the \textit{Local~MF}, respectively~\cite{Chen21}. While the primary focus of this paper is to design the joint network-wide multi-group multicast precoders at the BSs in a fully distributed fashion, we point out that the \textit{Centralized}, the \textit{Local~MMSE}, and the \textit{Local~MF} are also part of our contribution as they are tailored for the sum MSE minimization in the multi-group multicasting scenario.

\subsection{Centralized Precoding Design with Pilot-Aided Channel Estimation} \label{sec:problem_ref}

The practical implementation of the \textit{Centralized} requires the antenna-specific uplink channel estimation (see Section~\ref{sec:SM_est}) to enable the computation of the precoders in \eqref{eq:bsbf_mse} and the combiners in \eqref{eq:uebf} at the CPU. First, each BS~$b$ obtains $\{\hat \H_{b,k}\}_{k \in \setK}$ and forwards them to the CPU via backhaul signaling. Then, the CPU computes the aggregated precoders $\{\w_g\}_{g \in \setG}$ and the combiners $\{\v_k\}_{k \in \setK}$ via alternating optimization by replacing $\H_k$ with $\hat \H_k \triangleq [\hat \H_{1,k}^{\tran}, \ldots, \hat \H_{B,k}^{\tran}]^{\tran} \in \Compl^{B M \times N}$ in \eqref{eq:bsbf_mse} and \eqref{eq:uebf}, respectively. After convergence, the resulting BS-specific precoders are fed back to the corresponding BSs via backhaul signaling. Finally, the effective downlink channel estimation (see Section~\ref{sec:SM_est}) is carried out to allow each UE~$k$ to compute its (final) combiner as
\begin{align}\label{eq:rxmmse}
\v_{k}  = \big(\Y_k^{\dl} (\Y_k^{\dl})^{\herm}\big)^{-1} \Y_k^{\dl} \p_{g_{k}}^{\dl}.
\end{align}
Note that \eqref{eq:rxmmse} coincides with \eqref{eq:uebf} for perfect CSI, i.e., when $\tau^{\dl} \to \infty$. The implementation of the \textit{Centralized} is summarized in Algorithm~\ref{alg:cen}.

\smallskip

\begin{algorithm}[t!]
\footnotesize
\begin{spacing}{1.25}
\textbf{Data:} Pilots $\{\P_{k}^{\ul}\}_{k \in \setK}$ and $\{\p_{g}^{\dl}\}_{g \in \setG}$.
\begin{itemize}
\item[1)] \textbf{UL:} Each UE~$k$ transmits $\X_{k}^{\ul}$ in \eqref{eq:X_k_ul}; each BS~$b$ receives $\Y_{b}^{\ul}$ in \eqref{eq:Y_b_ul}.
\item[2)] Each BS~$b$ obtains $\{\hat \H_{b,k}\}_{k \in \setK}$ in \eqref{eq:H_bk_hat_orth} and forwards them to the CPU via backhaul signaling.
\end{itemize}
\textbf{Initialization:} Combiners $\{\v_{k}\}_{k \in \setK}$.\\
\textbf{Until} a predefined termination criterion is satisfied, \textbf{do:}
\begin{itemize}
\item[3)] The CPU computes the precoders $\{ \w_{g} \}_{g}$ as in \eqref{eq:bsbf} and the combiners $\{ \v_{k} \}_{k \in \setK}$ as in \eqref{eq:uebf} by replacing $\{ \H_{k} \}_{k \in \setK}$ with $\{ \hat \H_{k} \}_{k \in \setK}$.
\end{itemize}
\textbf{End}
\begin{itemize}
\item[4)] The CPU forwards the resulting BS-specific precoders to the corresponding BSs via backhaul signaling.
\item[5)] \textbf{DL:} Each BS~$b$ transmits $\X_{b}^{\dl}$ in \eqref{eq:X_b_dl}; each UE~$k$ receives $\Y_{k}^{\dl}$ in \eqref{eq:Y_k_dl}.
\item[6)] Each UE~$k$ computes its combiner $\v_{k}$ as in \eqref{eq:rxmmse}.
\end{itemize}
\end{spacing}
\caption{(\textit{Centralized})} \label{alg:cen}
\end{algorithm}

\begin{figure*}
\addtocounter{equation}{+3}
\begin{align} \label{eq:w_bg_*}
\Delta \w_{b,g}^{\star} & = -\bigg( \sum_{ k \in \setK}\omega_{ k}\H_{b, k}\v_{ k} \v_{ k}^{\herm} \H_{b, k} ^{\herm} + \lambda_{b}\I_M  \bigg)^{-1}  \bigg(\sum_{ k\in \setK_g}\omega_{k} \H_{b,k} \v_{k}  - \sum_{\bar b \in \setB} \sum_{ k \in \setK}\omega_{ k}\H_{b, k}\v_{ k} \v_{ k}^{\herm} \H_{\bar b, k} ^{\herm}\w_{\bar b,g}^{(i-1)} - \lambda_{b} \w_{b,g}^{(i-1)} \bigg)
\end{align}
\addtocounter{equation}{-3}
\hrulefill
\end{figure*}

\section{Distributed Precoding Design with Perfect CSI} \label{sec:distr-perf}

In this section, we describe the proposed distributed multi-group multicast precoding designs with perfect CSI and backhaul signaling for the CSI exchange. Their practical implementation with imperfect CSI and without any backhaul signaling for the CSI exchange is presented in Section~\ref{sec:distr-imperf}. The precoders are optimized locally at each BS by means of either best-response or gradient-based updates, as discussed in the following sections. Regardless of the computation of the precoders, each UE~$k$ computes its combiner as in \eqref{eq:uebf} with perfect CSI.

\subsection{Best-Response Distributed Precoding Design} \label{sec:distr-perf_BR}

In the best-response distributed precoding design, which is referred to in the following as the \textit{Distributed~BR}, the optimal $\w_{b,g}$ is obtained by setting $\nabla_{\w_{b,g}} \mathcal{L}_{\eqref{eq:EqvprobForHi}} \big( \{ \w_g, \lambda_b \} \big) =\0$, which yields
\addtocounter{equation}{-1}
\begin{align} \label{eq:dis_bsbf}
\w_{b,g} & = \bigg( \! \sum_{ k \in \setK}\omega_{k}\H_{b,k}\v_{ k} \v_{ k}^{\herm} \H_{b,k} ^{\herm} \! + \! \lambda_{b} \I_M \! \bigg)^{-1} \! \bigg( \! \sum_{ k\in \setK_g} \omega_{k} \H_{b,k} \v_{k} \nonumber \\
& \phantom{=} \ - \underbrace{\sum_{\bar b \ne b} \sum_{k \in \setK} \omega_{ k}\H_{b, k}\v_{ k} \v_{k}^{\herm} \H_{\bar b,  k}^{\herm}\w_{\bar b,g}}_{\scriptsize \triangleq \, \xib_{b,g}~\textnormal{(cross terms)}} \bigg).
\end{align}
The above precoder can be computed locally at BS~$b$ provided that $\xib_{b,g}$, which comprises group-specific cross terms from the other BSs, is known. To reconstruct $\xib_{b,g}$, BS~$b$ needs to obtain $\{\v_k^{\herm} \H_{\bar{b},k}^{\herm} \w_{\bar{b},g}\}_{k \in \setK}$ from each BS~$\bar{b} \neq b$ via backhaul signaling as in~\cite{Kal18}. In practice, each BS is required to share $G K$ complex scalars with the other BSs. In addition, the backhaul signaling introduces a delay that causes each BS to reconstruct the cross terms based on outdated CSI from the other BSs. As done in~\cite{Atz21}, we assume that such a delay consists of a single bi-directional training iteration. Hence, the cross terms $\xib_{b,g}$ at iteration~$i$ are given by $\xib_{b,g}^{(i)} \triangleq {\sum_{\bar b \ne b} \sum_{ k \in \setK}\omega_{ k}\H_{b, k}\v_{ k} \v_{k}^{\herm} \H_{\bar b, k} ^{\herm}\w_{\bar b,g}^{(i-1)}}$. 
With this information, all the BSs can compute their precoders simultaneously building on the parallel optimization framework~\cite{Scu14}, which uses best-response updates to ensure the convergence to a solution of the sum MSE minimization in \eqref{eq:EqvprobForHi}. Finally, the BS-specific precoder at iteration~$i$ is computed as
\begin{align} \label{eq:w_bk_i}
\w_{b,g}^{(i)} & = (1-\alpha_{\textnormal{\tiny{BR}}}) \w_{b,g}^{(i-1)} + \alpha_{\textnormal{\tiny{BR}}} \w_{b,g} \\
& = \w_{b,g}^{(i-1)} -\alpha_{\textnormal{\tiny{BR}}} \underbrace{(\w_{b,g}^{(i-1)}-\w_{b,g})}_{\triangleq \Delta\w_{b,g}^{\star}},
\end{align}
\addtocounter{equation}{+1}
where the step size $\alpha_{\textnormal{\tiny{BR}}} \in (0,1]$ strikes a balance between convergence speed and accuracy of the solution~\cite{Scu14}, and $\Delta \w_{b,g}^{\star}$ is obtained by replacing $\xib_{b,g}$ with $\xib_{b,g}^{(i)}$ in \eqref{eq:dis_bsbf} as shown in \eqref{eq:w_bg_*} at the top of the next page.

\begin{theorem} \label{th:disStepDis}
$\Delta \w_{b,g}^{\star}$ in \eqref{eq:w_bg_*} is a steepest descent direction for the sum MSE minimization in \eqref{eq:EqvprobForHi}.
\end{theorem}

\begin{IEEEproof}
Let us write the gradient of \eqref{eq:lagbsmse} with respect to $\w_{b,g}$ as
\begin{align} \label{eq:msegradient}
& \nabla_{\w_{b,g}} \mathcal{L}_{\eqref{eq:EqvprobForHi}} \big( \{\w_{g}, \lambda_b\} \big) \nonumber \\
& = - 2 \bigg( \sum_{k \in \setK_g} \omega_{k} \H_{b,k} \v_{k} - \sum_{\bar b \in \setB} \sum_{k \in \setK} \omega_{k} \H_{b,k} \v_{k} \v_{k}^{\herm} \H_{\bar b, k} ^{\herm}\w_{\bar b,g} \nonumber \\
& \phantom{=} \ - \lambda_{b} \w_{b,g} \bigg).
\end{align}
Furthermore, let us define $\C_b \triangleq 2 \big( \sum_{k \in \setK} \omega_{ k} \H_{b,k} \v_{k} \v_{ k}^{\herm} \H_{b, k}^{\herm} + \lambda_{b} \I_M \big) \in \Compl^{M \times M}$ and $\C \triangleq \blkdiag(\C_1,\ldots,\C_B) \in \Compl^{BM \times BM}$. Then, we simplify \eqref{eq:w_bg_*} as
\begin{align}
\label{eq:diPreSimp} \Delta \w_{b,g}^{\star} = \C_b^{-1} \nabla_{\w_{b,g}} \mathcal{L}_{\eqref{eq:EqvprobForHi}} \big( \{ \w_{g}, \lambda_b \} \big)
\end{align}
and, exploiting the fact that $\C^{-1} = \blkdiag(\C_1^{-1},\ldots,\C_B^{-1})$, we have
\begin{align} \label{eq:sdequ}
    \Delta \w_g^{\star} & \triangleq \big[ (\Delta \w_{1,g}^{\star})^{\tran}, \ldots, (\Delta \w_{B,g}^{\star})^{\tran} \big]^{\tran} \\
    & = \C^{-1}\nabla_{\w_{g}} \mathcal{L}_{\eqref{eq:EqvprobForHi}}\big(\{ \w_{g},\lambda_b\}\big).
\end{align}
Finally, we observe that $\Delta \w_g^{\star}$ in \eqref{eq:sdequ} is a steepest descent direction for the quadratic norm $\|\x\|_{\C} \triangleq (\x^{\herm} \C \x)^{\frac{1}{2}}$~\cite{Boy04}.  
\end{IEEEproof}

\begin{remark} \label{re:th1_step} \rm{
Theorem~\ref{th:disStepDis} states that, for a fixed set of combiners $\{\v_k\}_{k \in \setK}$, the \textit{Distributed~BR} solves the sum MSE minimization in \eqref{eq:EqvprobForHi} via a steepest descent method characterized by the quadratic norm $\|\x\|_{\C}$. Since $\C$ is a block-diagonal matrix with blocks $\{ \C_{b} \}_{b \in \setB}$, each BS~$b$ greedily aims to reduce its individual MSE by following the steepest descent direction for the quadratic norm $\|\x\|_{\C_{b}}$, whereas the convergence to a solution of the sum MSE minimization is guaranteed by a proper choice of $\alpha_{\textnormal{\tiny{BR}}}$. On the other hand, the centralized precoding design with best-response updates is obtained by replacing $\C$ with the Hessian of \eqref{eq:lagbsmse} in \eqref{eq:sdequ}, where the latter is a full matrix. Therefore, the \textit{Distributed~BR} is not equivalent to its centralized implementation and, as a consequence, may be characterized by slow convergence. This motivates the development of the gradient-based distributed precoding design in Section~\ref{sec:distr_perf_GB}. Lastly, we point out that the outdated CSI used to reconstruct the cross terms at each BS further slows down the convergence.}
\end{remark}

\begin{remark} \label{re:Sp_conv} \rm{
To speed up the convergence of the \textit{Distributed~BR}, we impose that, for a fixed set of combiners $\{\v_k\}_{k \in \setK}$, the BS-specific precoders $\{\w_{b,g}\}_{g \in \setG}$ are updated only once at each BS~$b$. In this respect, a sufficiently small $\alpha_{\textnormal{\tiny{BR}}}$ would ensure the monotonic (yet slow) convergence to a solution of the sum MSE minimization in \eqref{eq:EqvprobForHi} even with a single update of the precoders for a fixed set of combiners~\cite{Scu14}. However, considering a practical scenario where only a limited number of bi-directional training iterations is admissible, we disregard the strictly monotonic convergence and choose $\alpha_{\textnormal{\tiny{BR}}}$ to promote an aggressive reduction of the sum MSE objective during the first few iterations.}
\end{remark}

\subsection{Gradient-Based Distributed Precoding Design} \label{sec:distr_perf_GB}

The \textit{Distributed~BR} presented in Section~\ref{sec:distr-perf_BR} is not equivalent to its centralized implementation and may be thus characterized by slow convergence (see Remark~\ref{re:th1_step}). Hence, in this section, we propose a gradient-based distributed precoding design, which is referred to in the following as the \textit{Distributed~GB} and follows directly from its centralized implementation. In this method, the BS-specific precoders are first updated using the gradient of the sum MSE objective and then projected to meet the per-BS transmit power constraints. To this end, we write the gradient of the sum MSE objective (cf. \eqref{eq:MSE_k}) with respect to $\w_{b,g}$ as
\begin{align} \label{eq:BSUd}
\nabla_{\w_{b,g}} \bigg( \! \sum_{k \in \setK} \omega_k \mse_k \! \bigg) & = - 2 \bigg( \! \sum_{k \in \setK_g} \omega_{k} \H_{b, k} \v_{k} \nonumber \\
& \phantom{=} \ - \sum_{\bar b \in \setB}\sum_{k \in \setK} \omega_{k} \H_{b,k} \v_{k} \v_{k}^{\herm} \H_{\bar b, k}^{\herm} \w_{\bar b,g} \! \bigg).
\end{align}
Then, the corresponding gradient-based update can be expressed as
\begin{align}
\tilde{\w}_{b,g}^{(i)} & \triangleq \w_{b,g}^{(i-1)} - \alpha_{\textnormal{\tiny{GB}}} \nabla_{\w_{b,g}}\bigg(\sum_{k \in \setK} \omega_k \mse_k \bigg) \\
\label{eq:BSgraUp3} & = \w_{b,g}^{(i-1)} + 2 \alpha_{\textnormal{\tiny{GB}}} \bigg( \sum_{k \in \setK_g} \omega_{k} \H_{b, k} \v_{k} - \xib_{b,g}^{(i)} \nonumber \\
& \phantom{=} \ - \sum_{k \in \setK} \omega_{ k}\H_{b, k}\v_{ k} \v_{ k}^{\herm} \H_{b,k}^{\herm} \w_{b,g}^{(i-1)} \bigg),
\end{align}
where $\alpha_{\textnormal{\tiny{GB}}}$ is the step size. The above gradient-based update can be computed locally at BS~$b$ upon receiving the CSI from the other BSs (necessary to reconstruct the cross terms) via backhaul signaling. Finally, the BS-specific precoders at iteration~$i$ are obtained by projecting $\{\tilde{\w}_{b,g}^{(i)}\}_{g \in \setG}$ to meet the per-BS transmit power constraint, i.e.,
\begin{align} \label{eq:BSBfproj}
& [\w_{b,1}^{(i)}, \ldots, \w_{b,G}^{(i)}] \nonumber \\
& = \underset{\{\w_{b, \bar g}\}_{\bar g \in \setG}: \sum_{\bar g \in \setG}\| \w_{b,\bar g}\|^{2} \leq \rho_{\bs}}{\argmin} \sum_{\bar g \in \setG}\| \w_{b,\bar g} - \tilde{\w}_{b, \bar g}^{(i)} \|^2 \\
& = a_b [\tilde{\w}_{b,1}^{(i)}, \ldots, \tilde{\w}_{b,G}^{(i)}],
\end{align}
with $a_b = \sqrt{\frac{{\rho_{\bs}}}{{\sum_{\bar g \in \setG}\| \tilde{\w}_{b,\bar g}^{(i)}\|^{2}}}}$ if $\sum_{\bar g \in \setG}\| \hat \w_{b,\bar g}^{(i)}\|^{2} \geq \rho_{\bs}$ and $a_b = 1$ otherwise. Note that this approach can be easily extended to a unicasting scenario considering a single UE in each multicast group.

\begin{theorem} \label{th:grEqui}
The \textit{Distributed~GB} is equivalent to its centralized implementation.
\end{theorem}

\begin{IEEEproof}
Considering the centralized implementation, the gradient of the sum MSE objective (cf. \eqref{eq:MSE_k}) with respect to $\w_{g}$ is given by
\begin{align}
& \nabla_{\w_g} \bigg(\sum_{k \in \setK} \omega_k \mse_k \bigg) \nonumber \\
& = - 2 \bigg( \sum_{k \in \setK_g} \omega_k \H_k \v_k - \sum_{k \in \setK} \omega_k \H_k \v_k \v_k^{\herm} \H_k^{\herm} \w_g \bigg) \\
& \label{eq:gdEqu} = \begin{bmatrix}
\nabla_{\w_{1,g}} \big( \sum_{k \in \setK} \omega_k \mse_k \big) \\ \vdots \\ \nabla_{\w_{B,g}} \big( \sum_{k \in \setK} \omega_k \mse_k \big)
\end{bmatrix},
\end{align}
which corresponds to the concatenation of the gradients with respect to the BS-specific precoders (see \eqref{eq:BSUd}). As a consequence, the gradient-based update of $\w_{g}$ can be expressed as the concatenation of the gradient-based updates of the BS-specific precoders (see \eqref{eq:BSgraUp3}) at iteration~$i$. Then, the aggregated precoders at iteration~$i$ are obtained by projecting the aforementioned gradient-based updates to meet the per-BS transmit power constraints, i.e.,
\begin{align} \label{eq:Cenproj}
& [\w_{1}^{(i)}, \ldots, \w_{G}^{(i)} ] \nonumber \\
& = \underset{\{ \{\w_{b,\bar g }\}_{\bar g \in \setG} : \sum_{\bar g \in \setG} \| \w_{b, \bar g} \|^{2} \leq \rho_{\bs} \}_{b \in \setB}}{\argmin} \sum_{b \in \setB} \sum_{\bar g \in \setG} \| \w_{b, \bar g} - \tilde{\w}_{b, \bar g}^{(i)} \|^2 \\
 & = \begin{bmatrix} a_1 [\tilde{\w}_{1, 1}^{(i)}, \ldots, \tilde{\w}_{1, G}^{(i)}] \\
\vdots \\
a_B [\tilde{\w}_{B, 1}^{(i)}, \ldots, \tilde{\w}_{B, G}^{(i)}]
\end{bmatrix} .
\end{align}
Finally, we observe that the aggregated precoders in \eqref{eq:Cenproj} correspond to the concatenation of the BS-specific precoders in \eqref{eq:BSBfproj}.
\end{IEEEproof}

\begin{remark} \rm{
Theorem~\ref{th:grEqui} states that, for a fixed set of combiners $\{\v_k\}_{k \in \setK}$, the \textit{Distributed~GB} (where the BS-specific precoders $\{ \w_{b,g} \}_{g \in \setG}$ are optimized locally at each BS~$b$) solves the sum MSE minimization in \eqref{eq:EqvprobForHi} in the same way as its centralized implementation (where the aggregated precoders $\{ \w_{g} \}_{g \in \setG}$ are optimized at the CPU). Therefore, each BS directly targets to reduce the sum MSE rather than its individual MSE as in the \textit{Distributed~BR}. Moreover, the convergence to a solution of the sum MSE minimization is guaranteed by a proper choice of $\alpha_{\textnormal{\tiny{GB}}}$. Lastly, the comments in Remark~\ref{re:Sp_conv} on how to speed up the convergence of the \textit{Distributed~BR} also apply here.}
\end{remark}

\begin{figure*}
\begin{align} \label{eq:bsbflocal}
\addtocounter{equation}{+5}
\Delta \w_{b,g}^{\star} & \simeq - \big( \Y_{b}^{\ulA} \Omegab ({\Y_{b}^{\ulA}})^{\herm} \! + \! \tau^{\ulA}({\beta^{\ulA}}\lambda_{b} \! - \! \sigma_{\bs}^2)\I_M \big)^{-1} \! \bigg( \! \sqrt{\beta^{\ulA}} \sum_{k \in \setK_g} \omega_k\Y_{b}^{\ulA}\p^{\ulA}_k \! - \! \frac{{\tau^{\ulA}\beta^{\ulA}}}{\tau^{\dl}\sqrt{\beta^{\ulC}}}\Y_{b}^{\ulC}(\p^{\dl}_g)^{\herm} \! - \! {{\beta^{\ulA}}} \tau^{\ulA}\lambda_b \w_{b,g}^{(i-1)} \! \bigg)
\end{align}
\hrulefill
\begin{align}\label{eq:bsbflocalgs}
\Delta \w_{b,g}^{\star} & \simeq - \big( \Y_{b}^{\ulB} (\Y_{b}^{\ulB})^{\herm} + \tau^{\ulB}(\beta^{\ulB} \lambda_{b} - \sigma_{\bs}^2) \I_M \big)^{-1} \bigg( \sqrt{\beta^{\ulB}} \Y_{b}^{\ulB} \p^{\ulB}_g - \frac{\beta^{\ulB} \tau^{\ulB}}{\sqrt{\beta^{\ulC}} \tau^{\dl}} \Y_{b}^{\ulC} (\p^{\dl}_g)^{\herm} - {\beta^{\ulB}} \tau^{\ulB} \lambda_b \w_{b,g}^{(i-1)} \bigg)
\end{align}
\addtocounter{equation}{-7}
\hrulefill
\end{figure*}

\section{Distributed Precoding Design with Pilot-Aided Channel Estimation} \label{sec:distr-imperf}

In this section, we describe the practical implementation of the proposed distributed multi-group multicast precoding designs with imperfect CSI and without any backhaul signaling for CSI exchange. We recall that the local computation of the precoders at each BS in \eqref{eq:dis_bsbf} relies on group-specific cross terms from the other BSs. To avoid the resulting CSI exchange via backhaul signaling, we adopt an OTA signaling scheme similar to that proposed in our previous work on distributed precoding design for cell-free massive MIMO unicasting~\cite{Atz21}. Therein, we introduced a UE-specific OTA uplink training resource to eliminate the need for backhaul signaling to exchange the UE-specific CSI. In this paper, we propose a new group-specific OTA uplink training resource tailored for the multi-group multicasting scenario, which eliminates the need for backhaul signaling to exchange the group-specific CSI.

\smallskip

\textit{\textbf{New group-specific OTA uplink training resource (UL-3).}} To reconstruct the cross terms $\xib_{b,g}$ locally at BS~$b$, each UE~$k$ transmits $\Y_{k}^{\dl}$ in \eqref{eq:Y_k_dl} after precoding it with $\omega_{k} \v_{k} \v_{k}^{\herm}$, i.e.,
\begin{align} \label{eq:Disultx3}
\X_{k}^{\ulC} \triangleq \sqrt{\beta^{\ulC}} \omega_{k} \v_{k} \v_{k}^{\herm} \Y_{k}^{\dl} \in \Compl^{N \times \tau^{\dl}},
\end{align}
where the power scaling factor $\sqrt{\beta^{\ulC}}$ (equal for all the UEs) ensures that $\X_{k}^{\ulC}$ complies with the per-UE transmit power constraint. We observe that \eqref{eq:Disultx3} contains the group-specific effective downlink channels between all the BSs and UE~$k$, and we recall that $\Y_{k}^{\dl}$ is obtained by means of group-specific pilots (see Section~\ref{sec:SM_est}). Therefore, this new group-specific OTA uplink training resource generates the same training overhead as the effective downlink channel estimation, which depends on $G$ rather than $K$ as in the unicasting scenario. Then, the signal received at BS~$b$ is given by
\begin{align}
\Y_{b}^{\ulC} & \triangleq \sum_{k \in \setK} \omega_{k} \H_{b,k} \v_{k} \v_{k}^{\herm} \Y_{k}^{\dl} + \Z_{k}^{\ulC} \\
\label{eq:Disulrx3} & = \sqrt{\beta^{\ulC}} \sum_{k \in \setK} \omega_{k} \H_{b,k} \v_{k}\v_{k}^{\herm} \bigg( \sum_{g \in \setG} \H_{k}^{\herm} \w_{g} (\p^{\dl}_{g})^{\herm} \nonumber \\
& \phantom{=} \ + \Z_k^{\dl} \bigg) + \Z_{k}^{\ulC} \in \Compl^{M \times \tau^{\dl}},
\end{align}
where $\Z_{b}^{\ulC}$ $ \in \Compl^{M \times \tau^{\dl}}$ is the AWGN with i.i.d. $\setC \setN (0, \sigma_{\bs}^{2})$ elements. Finally, the LS estimate of $\xib_{b,g}$ is   
\begin{align} \label{eq:cr_re}
\hat{\xib}_{b,g} & \triangleq \frac{1}{\tau^{\dl} \sqrt{\beta^{\ulC}}} \Y_{b}^{\ulC}\p^{\dl}_g \\ &
= \xib_{b,g} + \frac{1}{\tau^{\dl}} \bigg(\sum_{\bar g \ne g} \xib_{b,\bar g} (\p^{\dl}_{\bar g})^\herm + \sum_{k \in \setK} \omega_{k} \H_{b,k} \v_{k} \v_{k}^{\herm} \Z_k^{\dl} \nonumber \\
&\phantom{=} \ + \frac{1}{\sqrt{\beta^{\ulC}}} \Z_{k}^{\ulC} \bigg) \p^{\dl}_g.
\end{align}

\smallskip

Building on the new group-specific OTA uplink training resource, the precoders are optimized locally at each BS by means of either best-response updates (based on both UE- and group-specific pilots or group-specific pilots only) or gradient-based updates (based on group-specific pilots), as discussed in the following sections. Regardless of the computation of the precoders, each UE~$k$ computes its combiner as in \eqref{eq:rxmmse} with imperfect CSI.

\subsection{Best-Response Distributed Precoding Design with UE- and Group-Specific Pilots} \label{sec:distr-imperf_BR}

The practical implementation of the \textit{Distributed~BR} requires, at each bi-directional training iteration, the UE-specific effective uplink channel estimation and the effective downlink channel estimation (see Section~\ref{sec:SM_est}) together with the new group-specific OTA uplink training resource (see Section~\ref{sec:distr-imperf}). In this setting, $\Y_{b}^{\ulA}$ in \eqref{eq:Y_b_ul1} and $\Y_{b}^{\ulC}$ in \eqref{eq:Disulrx3} are suitably combined to reconstruct $\Delta \w_{b,g}^{\star}$ in \eqref{eq:w_bg_*} as shown in \eqref{eq:bsbflocal} at the top of the next page, which is used to compute the BS-specific precoder in \eqref{eq:w_bk_i}. Note that \eqref{eq:bsbflocal} becomes equal to \eqref{eq:w_bg_*} with perfect CSI, i.e., when $\tau^{\ulA} \to \infty$ and $\tau^{\dl} \to \infty$. If pilot contamination is to be avoided entirely, the \textit{Distributed~BR} requires a minimum of $K+G$ orthogonal pilots, i.e., $K$ orthogonal pilots to obtain $\Y_{b}^{\ulA}$ in \eqref{eq:Y_b_ul1} and $G$ orthogonal pilots to obtain $\Y_{b}^{\ulC}$ in \eqref{eq:Disulrx3}, in each uplink training instance. The implementation of the \textit{Distributed~BR} is summarized in Algorithm~\ref{alg:disNW}.

\begin{algorithm}[t!]
\footnotesize
\begin{spacing}{1.25}
\textbf{Data:} Pilots $\{\p^{\ulA}_{k}\}_{k \in \setK}$  and $\{\p^{\dl}_{g}\}_{g \in \setG}$.\\
\textbf{Initialization:} Combiners $\{\v_{k}\}_{k \in \setK}$.\\
\textbf{Until} a predefined termination criterion is satisfied, \textbf{do:}
\begin{itemize}
\item[1)] \textbf{UL-1:} Each UE~$k$ transmits $\X_{k}^{\ulA}$ in \eqref{eq:X_k_ul1}; each BS~$b$ receives $\Y_{b}^{\ulA}$ in \eqref{eq:Y_b_ul1}.
\item[2)] \textbf{UL-3:} Each UE~$k$ transmits $\X_{k}^{\ulC}$ in \eqref{eq:Disultx3}; each BS~$b$ receives $\Y_{b}^{\ulC}$ in \eqref{eq:Disulrx3}.
\item[3)] Each BS~$b$ reconstructs $\{ \Delta \w_{b,g}^{\star} \}_{g \in \setG}$ as in \eqref{eq:bsbflocal} and computes the precoders $\{ \w_{b,g} \}_{g \in \setG}$ as in \eqref{eq:w_bk_i}.
\item[4)] \textbf{DL:} Each BS~$b$ transmits $\X_{b}^{\dl}$ in \eqref{eq:X_b_dl}; each UE~$k$ receives $\Y_{k}^{\dl}$ in \eqref{eq:Y_k_dl}.
\item[5)] Each UE~$k$ computes its combiner $\v_{k}$ as in \eqref{eq:rxmmse}.
\end{itemize}
\textbf{End}
\caption{(\textit{Distributed~BR})} \label{alg:disNW}
\end{spacing}
\end{algorithm}

\subsection{Best-Response Distributed Precoding Design with Group-Specific Pilots} \label{sec:distr-imperf_BR-GS}

The practical implementation of the \textit{Distributed~BR} described in Section~\ref{sec:distr-imperf_BR} relies on the UE-specific effective uplink channel estimation, which requires a minimum of $K$ orthogonal pilots in each uplink training instance to avoid pilot contamination. Hence, to reduce the training overhead, we propose a best-response distributed precoding design based solely on group-specific pilots, which is referred to in the following as the \textit{Distributed~BR-GS}. This method is obtained by replacing the UE-specific effective uplink channel estimation with its group-specific counterpart (see Section~\ref{sec:SM_est}). Consequently, if pilot contamination is to be avoided entirely, the \textit{Distributed~BR-GS} requires a minimum of $2 G < K + G$ orthogonal pilots, i.e., $G$ orthogonal pilots to obtain $\Y_{b}^{\ulB}$ in \eqref{eq:Y_b_ul2} and $G$ orthogonal pilots to obtain $\Y_{b}^{\ulC}$ in \eqref{eq:Disulrx3}, in each uplink training instance. In this setting, assuming $\omega_k=\omega,~\forall k \in \setK$, $\Y_{b}^{\ulB}$ in \eqref{eq:Y_b_ul2} and $\Y_{b}^{\ulC}$ in \eqref{eq:Disulrx3} are suitably combined to reconstruct $\Delta \w_{b,g}^{\star}$ in \eqref{eq:w_bg_*} as shown in \eqref{eq:bsbflocalgs} at the top of the page, which is used to compute the BS-specific precoder in \eqref{eq:w_bk_i}. To understand the convergence behavior of the \textit{Distributed~BR-GS}, let us assume for a moment that perfect CSI is available at BS~$b$, i.e., $\tau^{\ulB} \to \infty$ and $\tau^{\dl} \to \infty$. In this case, we have
\addtocounter{equation}{+2}
\begin{align} \label{eq:diGSPreSimp}
\Delta \w_{b,g}^{\star} & \simeq \D_b^{-1} \nabla_{\w_{b,g}} \mathcal{L}_{\eqref{eq:EqvprobForHi}}\big(\{ \w_{g}, \lambda_b\}\big)
\end{align}
with $\nabla_{\w_{b,g}} \mathcal{L}_{\eqref{eq:EqvprobForHi}} \big( \{ \w_g,\lambda_b \} \big)$ given in \eqref{eq:msegradient} and
\begin{align}
\D_b & \triangleq 2 \bigg( \sum_{k \in \setK} \H_{b, k} \v_{ k} \v_{ k}^{\herm} \H_{b, k} ^{\herm}  \nonumber \\ 
& \phantom{=} \ + \underbrace{\sum_{\bar{g} \in \setG}  \bigg( \sum_{k \in \setK_{\bar{g}}} \H_{b,k} \v_{k} \bigg) \bigg( \sum_{\bar k \in \setK_{\bar{g}} \setminus \{k\}} \v_{\bar k}^{\herm} \H_{b,\bar k}^{\herm} \bigg)}_{\textnormal{extra interference}} + \lambda_{b} \I_M \bigg).
\end{align}
We observe that \eqref{eq:diGSPreSimp} includes an extra interference term with respect to \eqref{eq:w_bg_*}, which arises from reconstructing the local interference covariance matrix based solely on group-specific CSI (i.e., $\Y_{b}^{\ulB}$ in \eqref{eq:Y_b_ul2}) rather than UE-specific CSI (i.e., $\Y_{b}^{\ulA}$ in \eqref{eq:Y_b_ul1}) as in \eqref{eq:bsbflocal} for the \textit{Distributed~BR}. The implementation of the \textit{Distributed~BR-GS} is summarized in Algorithm~\ref{alg:disGS}.

\begin{algorithm}[t!]
\footnotesize
\begin{spacing}{1.25}
\textbf{Data:} Pilots $\{\p^{\ulB}_{g}\}_{g \in \setG}$  and $\{\p^{\dl}_{g}\}_{g \in \setG}$. \\
\textbf{Initialization:} Combiners $\{\v_{k}\}_{k \in \setK}$.\\
\textbf{Until} a predefined termination criterion is satisfied, \textbf{do:}
\begin{itemize}
\item[1)] \textbf{UL-2:} Each UE~$k$ transmits $\X_{k}^{\ulB}$ in \eqref{eq:X_g_ul2}; each BS~$b$ receives $\Y_{b}^{\ulB}$ as in \eqref{eq:Y_b_ul2}.
\item[2)] \textbf{UL-3:} Each UE~$k$ transmits $\X_{k}^{\ulC}$ in \eqref{eq:Disultx3}; each BS~$b$ receives $\Y_{b}^{\ulC}$ as in \eqref{eq:Disulrx3}.
\item[3)] Each BS~$b$ reconstructs $\{ \Delta \w_{b,g}^{\star} \}_{g \in \setG}$ as in \eqref{eq:bsbflocalgs} and computes the precoders $\{ \w_{b,g} \}_{g \in \setG}$ as in \eqref{eq:w_bk_i}.
\item[4)] \textbf{DL:} Each BS~$b$ transmits $\X_{b}^{\dl}$ in \eqref{eq:X_b_dl}; each UE~$k$ receives $\Y_{k}^{\dl}$ in \eqref{eq:Y_k_dl}.
\item[5)] Each UE~$k$ computes its combiner $\v_{k}$ as in \eqref{eq:rxmmse}.
\end{itemize}
\textbf{End}
\caption{(\textit{Distributed~BR-GS})} \label{alg:disGS}
\end{spacing}
\end{algorithm}

\begin{theorem} \label{th:disGsud}
$\Delta \w_{b,g}^{\star}$ in \eqref{eq:diGSPreSimp} is a steepest descent direction for the sum MSE minimization in \eqref{eq:EqvprobForHi}.
\end{theorem}

\begin{IEEEproof}
The proof follows similar steps to the proof of Theorem~\ref{th:disStepDis} and is thus omitted. 
\end{IEEEproof}

\begin{remark} \rm{
Following similar arguments to Remark~\ref{re:th1_step}, Theorem~\ref{th:disGsud} states that, for a fixed set of combiners $\{\v_k\}_{k \in \setK}$, the \textit{Distributed~BR-GS} solves the sum MSE minimization in \eqref{eq:EqvprobForHi} via a steepest descent method characterized by the quadratic norm $\| \x \|_{\D}$, with $\D \triangleq \blkdiag (\D_{1}, \ldots, \D_{B}) \in \Compl^{B M \times B M}$. Due to the extra interference term in \eqref{eq:diGSPreSimp}, the \textit{Distributed~BR-GS} may be characterized by slower convergence than the \textit{Distributed~BR}. Nonetheless, as shown in Section~\ref{sec:num}, this drawback may be well compensated by the reduced training overhead, especially for small resource blocks. Hence, the \textit{Distributed~BR-GS} may outperform the \textit{Distributed~BR} in terms of effective sum-group rate. Lastly, the comments in Remark~\ref{re:Sp_conv} on how to speed up the convergence of the \textit{Distributed~BR} also apply here.}
\end{remark}

\subsection{Gradient-Based Distributed Precoding Design with Group-Specific Pilots} \label{sec:distr-imperf_GB}

The practical implementation of the \textit{Distributed~GB} requires, at each bi-directional training iteration, the group-specific effective uplink channel estimation and the effective downlink channel estimation (see Section~\ref{sec:SM_est}) together with the new group-specific OTA uplink training resource (see Section~\ref{sec:distr-imperf}). In this setting, $\Y_{b}^{\ulB}$ in \eqref{eq:Y_b_ul2} and $\Y_{b}^{\ulC}$ in \eqref{eq:Disulrx3} are suitably combined to reconstruct $\nabla_{\w_{b,g}} \big( \sum_{k \in \setK} \omega_k \mse_k \big)$ in \eqref{eq:BSUd} as
\begin{align} \label{eq:bfgdstep}
\nabla_{\w_{b,g}} \bigg( \sum_{k \in \setK} \omega_k \mse_k \bigg) & \simeq \frac{2}{\tau^{\dl}\sqrt{\beta^{\ulC}}}\Y_{b}^{\ulC}\p^{\dl}_g \nonumber \\
&\phantom{=} \ - \frac{2}{\tau^{\ulB}\sqrt{\beta^{\ulB}}}\Y_{b}^{\ulB}\p^{\ulB}_g,
\end{align}
which is used to compute the corresponding gradient-based update in \eqref{eq:BSgraUp3}. Note that \eqref{eq:bfgdstep} becomes equal to \eqref{eq:BSUd} with perfect CSI, i.e., when $\tau^{\ulA} \to \infty$ and $\tau^{\dl} \to \infty$. Finally, the BS-specific precoders are obtained by projecting the gradient-based updates to meet the per-BS transmit power constraint as in \eqref{eq:BSBfproj}. Remarkably, the \textit{Distributed~GB} can be implemented based solely on group-specific pilots. Consequently, if pilot contamination is to be avoided entirely, the \textit{Distributed~GB} requires a minimum of $2 G$ orthogonal pilots in each uplink training instance (as the \textit{Distributed~BR-GS}). Another significant advantage of the \textit{Distributed~GB} is that the computation of the precoders does not involve any matrix inversion, which yields a reduced computational complexity with respect to the \textit{Distributed~BR} and the \textit{Distributed~BR-GS}. The implementation of the \textit{Distributed~GB} is summarized in Algorithm~\ref{alg:disGd}.

\begin{algorithm}[t!]
\footnotesize
\begin{spacing}{1.25}
\textbf{Data:} Pilots $\{\p^{\ulB}_{g}\}_{g \in \setG}$ and $\{\p^{\dl}_{g}\}_{g \in \setG}$. \\
\textbf{Initialization:} Combiners $\{\v_{k}\}_{k \in \setK}$.\\
\textbf{Until} a predefined termination criterion is satisfied, \textbf{do:}
\begin{itemize}
\item[1)] \textbf{UL-2:} Each UE~$k$ transmits $\X_{k}^{\ulB}$ in \eqref{eq:X_g_ul2}; each BS~$b$ receives $\Y_{b}^{\ulB}$ as in \eqref{eq:Y_b_ul2}.
\item[2)] \textbf{UL-3:} Each UE~$k$ transmits $\X_{k}^{\ulC}$ in \eqref{eq:Disultx3}; each BS~$b$ receives $\Y_{b}^{\ulC}$ as in \eqref{eq:Disulrx3}.
\item[3)] Each BS~$b$ reconstructs the gradient of the sum MSE objective with respect to $\{ \w_{b,g} \}_{g \in \setG}$ as in \eqref{eq:bfgdstep}, computes the corresponding gradient-based updates in \eqref{eq:BSgraUp3}, and projects them as in \eqref{eq:BSBfproj}.
\item[4)] \textbf{DL:} Each BS~$b$ transmits $\X_{b}^{\dl}$ in \eqref{eq:X_b_dl}; each UE~$k$ receives $\Y_{k}^{\dl}$ in \eqref{eq:Y_k_dl}.
\item[5)] Each UE~$k$ computes its combiner $\v_{k}$ as in \eqref{eq:rxmmse}.
\end{itemize}
\textbf{End}
\caption{(\textit{Distributed~GB})} \label{alg:disGd}
\end{spacing}
\end{algorithm}

\begin{table*}[t!]
\centering
\footnotesize
\begin{tabular}{|c|c|c|c|c|c|c|}
    \hline
    \textbf{Algorithm}  & \textbf{UL} ($\tau^{\ul}$) & \textbf{UL-1} ($\tau^{\ulA}$) & \textbf{UL-2} ($\tau^{\ulB}$) & \textbf{UL-3} ($\tau^{\dl}$) & \textbf{DL} ($\tau^{\dl}$) & \textbf{Total} \\
    \hline
    \textit{Centralized} (reference) & $ KN $ & -- & -- & -- & $G$ & $K N + G$ \\
    \hline
    \textit{Local~MMSE} (reference) & -- & $K$ & -- & -- & $G$ & $(K + G) I$ \\
    \hline
    \textit{Local~MF} (reference) & -- & -- & $G$ & -- & $ G$  & $2 G I$ \\
    \hline
    \textit{Distributed~BR} (proposed) & -- & $ K $ & -- & $G$ & $G$ & $(K + 2G) I$ \\
    \hline
    \textit{Distributed~BR-GS} (proposed) & -- & -- & $G$ & $G$ & $G$ & $3 G I$ \\
    \hline
    \textit{Distributed~GB} (proposed) & -- & -- & $G$ & $G$ & $G$ & $3 G I$ \\
    \hline
\end{tabular}
\caption{Minimum number of pilot symbols necessary for the iterative bi-directional training without pilot contamination in the proposed and reference precoding schemes, where $I$ denotes the total number of bi-directional training iterations (recall that the \textit{Centralized} requires a single uplink-downlink training iteration).}
\label{tab:pilots}
\end{table*}

\begin{table*}[t!]
\centering
\setlength{\tabcolsep}{4pt}
\footnotesize
\begin{tabular}{|c|c|c|c|c|c|c|}
    \hline
    \textbf{Algorithm} & \textit{Centralized} & \textit{Local~MMSE} & \textit{Local~MF} & \textit{Distributed~BR} & \textit{Distributed~BR-GS} & \textit{Distributed~GB} \\
    \hline
    \textbf{Complexity} & $\mathcal{O}( B^3M^3)$ & $\mathcal{O}(\delta B M (M^2+K^2))$ & $\mathcal{O}(\delta  B M G^2)$ & $\mathcal{O}(\delta  B M(M^2+K^2))$ & $\mathcal{O}(\delta  B M(M^2+G^2))$ & $\mathcal{O}(\delta  B MG^2)$ \\
    \hline
\end{tabular}
\caption{Computational complexity for each bi-directional training iteration of the proposed and reference precoding schemes, where $\delta$ denotes the number of bi-section steps at each iteration (recall that the \textit{Centralized} requires a single uplink-downlink training iteration).}
\label{tab:comp_comp}
\end{table*}

\subsection{Training Overhead}

The practical implementation of the proposed distributed precoding designs requires, at each bi-directional training iteration, the UE- or group-specific effective uplink channel estimation and the effective downlink channel estimation (see Section~\ref{sec:SM_est}). In addition, it also relies on the new group-specific OTA uplink training resource (see Section~\ref{sec:distr-imperf}), which eliminates the need for backhaul signaling to exchange the group-specific CSI. Consequently, in each uplink training instance, each UE~$k$ transmits $\X_k^{\ulA}$ in \eqref{eq:X_k_ul1} or $\X_k^{\ulB}$ in \eqref{eq:X_g_ul2} together with $\X_k^{\ulC}$ in \eqref{eq:Disultx3}. Similarly, in each downlink training instance, each BS~$b$ transmits $\X_b^{\dl}$ in \eqref{eq:X_b_dl}. In principle, the iterative bi-directional training comprising the above signaling can be integrated into the flexible 3GPP 5G NR frame/slot structure, as discussed in~\cite{Tol19,Atz21}. Table~\ref{tab:pilots} shows the minimum number of orthogonal pilots (and thus the minimum number of pilot symbols) necessary for the iterative bi-directional training without pilot contamination in the proposed and reference precoding schemes.

\begin{remark}\rm{
The \textit{Distributed~GB}, if implemented via backhaul signaling for the CSI exchange similarly to~\cite{Kal18}, would still require the UE-specific effective uplink channel estimation (see Section~\ref{sec:SM_est}) and would generate the same backhaul signaling overhead as the \textit{Distributed~BR} described in Section~\ref{sec:distr-perf_BR}. In fact, reconstructing the cross terms $\xib_{b,g}$ in \eqref{eq:dis_bsbf} at BS~$b$ is not possible with group-specific CSI exchange. On the other hand, adopting iterative bi-directional training with the new group-specific OTA uplink training resource allows to implement the \textit{Distributed~GB} (and the \textit{Distributed~BR-GS}) with reduced training overhead with respect to the \textit{Distributed~BR}.}
\end{remark}

\subsection{Computational Complexity}

Based on the minimum number of pilot symbols specified in Table~\ref{tab:pilots}, Table~\ref{tab:comp_comp} presents the computational complexity for each bi-directional training iteration of the proposed and reference precoding schemes. The computational complexity mainly arises from matrix multiplications and inversions in the computation of the precoders. Notably, the \textit{Local~MF} and the \textit{Distributed~GB} exhibit remarkably low computational complexity compared with the other methods. Additionally, the \textit{Distributed~BR-GS} is less complex than the \textit{Distributed~BR} as the former relies solely on group-specific pilots. Among all the considered methods, the \textit{Centralized} entails the highest computational complexity.

\section{Numerical Results and Discussion} \label{sec:num}

In this section, we compare the performance of the proposed distributed multi-group multicast precoding designs presented in Section~\ref{sec:distr-imperf}, i.e., the \textit{Distributed~BR} (Algorithm~\ref{alg:disNW}), the \textit{Distributed~BR-GS} (Algorithm~\ref{alg:disGS}), and the \textit{Distributed~GB} (Algorithm~\ref{alg:disGd}), with that of the reference precoding schemes described in Section~\ref{sec:problem_ref} and Appendix~\ref{app:A3}, i.e., the \textit{Centralized} (Algorithm~\ref{alg:cen}), the \textit{Local~MMSE}, and the \textit{Local~MF}. Unless otherwise stated, the simulation setup comprises the following parameters. $B=25$~BSs, each equipped with $M=8$~antennas, are placed on a square grid with a distance of $100$~m between neighboring BSs. $K=32$~UEs, each equipped with $N = 2$~antennas, are uniformly distributed across the square grid. The UEs are divided into $G=8$ multicast groups, each consisting of $4$ randomly selected UEs.\footnote{If the multicasting services demand the UEs to be grouped based on similar geographical locations, the interference among the multicast groups could be mitigated more effectively, thus yielding better performance with respect to the considered random UE grouping.} Assuming uncorrelated Rayleigh fading, the entries of $\H_{b,k}$ are i.i.d. $\setC \setN (0, \delta_{b,k})$ random variables, where $\delta_{b,k} \triangleq -48-30\log_{10} (d_{b,k})$~[dB] is the large-scale fading coefficient and $d_{b,k}$ is the distance between BS~$b$ and UE~$k$.\footnote{The simulation results would be very similar with correlated channel models such as the one-ring model~\cite{Abd02}.} The maximum transmit power for both data and the pilot transmission is $\rho_{\bs} = 30$~dBm at the BSs and $\rho_{\ue} = 20$~dBm at the UEs. The AWGN power at the BSs and at the UEs is fixed to $\sigma_{\bs}^{2} = \sigma_{\ue}^{2} = -95$~dBm. As a performance metric, we evaluate the sum-group rate in \eqref{eq:R} averaged over $10^3$ independent channel realizations and UE drops. In all the algorithms, the combiners at the UEs are initialized with random vectors and the step sizes are appropriately chosen to promote an aggressive reduction of the sum MSE objective during the first few iterations.

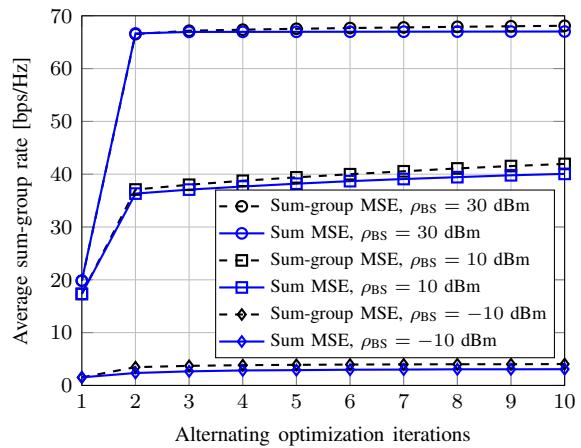
\begin{figure}[t]
\centering
\begin{tikzpicture}

\begin{axis}[
	width=8cm,
	height=6.5cm,
	xmin=1, xmax=10,
	ymin=0, ymax=70,
    xlabel={Alternating optimization iterations},
    ylabel={Average sum-group rate [bps/Hz]},
	xtick={1,2,3,4,5,6,7,8,9,10},
    ytick={0,10,20,30,40,50,60,70},
    xlabel near ticks,
	ylabel near ticks,
    x label style={font=\footnotesize},
	y label style={font=\footnotesize},
    ticklabel style={font=\footnotesize},
    legend style={at={(0.98,0.53)}, anchor=north east},
	legend style={font=\scriptsize, inner sep=1pt, fill opacity=0.75, draw opacity=1, text opacity=1},
	legend cell align=left,
	grid=both,
]

\addplot[thick, black,mark=o, dashed, mark options={solid}]
table[x=xin, y=grpMSEp30dBm, col sep=comma] 
{Figdata/fig2data.txt};
\addlegendentry{Sum-group MSE, $\rho_{\bs} = 30$~dBm};

\addplot[thick, blue,  mark=o]
table[x=xin, y=sumMSEp30dBm, col sep=comma] 
{Figdata/fig2data.txt};
\addlegendentry{Sum MSE, $\rho_{\bs} = 30$~dBm};

\addplot[thick, black,mark=square, dashed, mark options={solid}]
table[x=xin, y=grpMSEp10dBm, col sep=comma] 
{Figdata/fig2data.txt};
\addlegendentry{Sum-group MSE, $\rho_{\bs} = 10$~dBm};

\addplot[thick, mark=square, blue]
table[x=xin, y=sumMSEp10dBm, col sep=comma] 
{Figdata/fig2data.txt};
\addlegendentry{Sum MSE, $\rho_{\bs} = 10$~dBm};

\addplot[thick, black, mark=diamond, dashed, mark options={solid}]
table[x=xin, y=grpMSEm10dBm, col sep=comma] 
{Figdata/fig2data.txt};
\addlegendentry{Sum-group MSE, $\rho_{\bs} = -10$~dBm};

\addplot[thick, blue, mark=diamond]
table[x=xin, y=sumMSEm10dBm, col sep=comma] 
{Figdata/fig2data.txt};
\addlegendentry{Sum MSE, $\rho_{\bs} = -10$~dBm};

\end{axis}

\end{tikzpicture}
\caption{Average sum-group rate resulting from the sum-group MSE minimization and the sum MSE minimization versus number of alternating optimization iterations for different values of $\rho_{\bs}$.}
\label{fig:gmseVsSmse}
\end{figure}

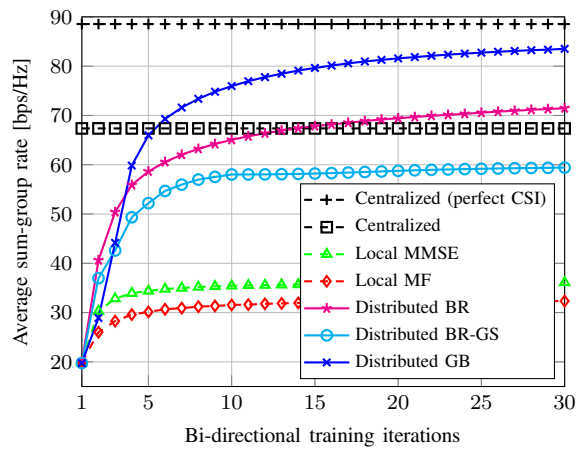
\begin{figure}[t]
\centering
\begin{tikzpicture}

\begin{axis}[
	width=8cm,
	height=6.5cm,
	xmin=1, xmax=30,
	ymin=15, ymax=90,
    xlabel={Bi-directional training iterations},
    ylabel={Average sum-group rate [bps/Hz]},
    xtick={1,5,10,15,20,25,30},	
    ytick={20,30,40,50,60,70,80,90},
    xlabel near ticks,
	ylabel near ticks,
    x label style={font=\footnotesize},
	y label style={font=\footnotesize},
    ticklabel style={font=\footnotesize},
    legend style={at={(0.98,0.02)}, anchor=south east},
	legend style={font=\scriptsize, inner sep=1pt, fill opacity=0.75, draw opacity=1, text opacity=1},
	legend cell align=left,
	grid=both,
]

\addplot[thick, black,mark=+, dashed, mark options={solid}]
table[x=xin, y=cenPer, col sep=comma] 
{Figdata/fig3data.txt};
\addlegendentry{Centralized (perfect CSI)};

\addplot[thick,black,mark=square, dashed, mark options={solid}]
table[x=xin, y=cen, col sep=comma] 
{Figdata/fig3data.txt};
\addlegendentry{Centralized};

\addplot[thick, green,mark=triangle, dashed, mark options={solid}]
table[x=xin, y=locMMSE, col sep=comma] 
{Figdata/fig3data.txt};
\addlegendentry{Local MMSE};

\addplot[thick,  red, mark=diamond, dashed, mark options={solid}]
table[x=xin, y=locMF, col sep=comma] 
{Figdata/fig3data.txt};
\addlegendentry{Local MF};

\addplot[thick, magenta, mark=star, mark options={solid}]
table[x=xin, y=disBR, col sep=comma] 
{Figdata/fig3data.txt};
\addlegendentry{Distributed BR};

\addplot[thick, cyan,mark=o, mark options={solid}]
table[x=xin, y=disBR-GS, col sep=comma] 
{Figdata/fig3data.txt};
\addlegendentry{Distributed BR-GS};

\addplot[thick, blue, mark=x]
table[x=xin, y=disGB, col sep=comma] 
{Figdata/fig3data.txt};
\addlegendentry{Distributed GB};

\end{axis}

\end{tikzpicture}
\caption{Average sum-group rate versus number of bi-directional training iterations.}
\label{fig:rateVsItr}
\end{figure}

We begin by validating Proposition~\ref{pro:1} considering a centralized implementation. Figure~\ref{fig:gmseVsSmse} compares the average sum-group rate resulting from the sum-group MSE minimization (see Section~\ref{sec:problem_sum-groupMSE}) and the sum MSE minimization (see Section~\ref{sec:problem_sumMSE}) for different values of $\rho_{\bs}$. We observe that, as the SNR increases, the gap between the two curves does not increase. Therefore, at high SNR, the sum-group rate obtained with the sum MSE minimization closely approximates the one resulting from the sum-group MSE minimization.

Figure~\ref{fig:rateVsItr} illustrates the average sum-group rate as a function of the number of bi-directional training iterations, where the \textit{Centralized} with perfect CSI is also included as an upper bound. The proposed distributed precoding designs greatly outperform the local precoding designs. During the first few iterations, the \textit{Distributed~BR} and the \textit{Distributed~BR-GS} are superior to the \textit{Distributed~GB}. Indeed, in the distributed precoding designs with best-response updates, each BS greedily aims to reduce its individual MSE by exploiting its local interference covariance matrix, yielding a slower convergence to a solution of the sum MSE minimization. On the other hand, the \textit{Distributed~GB} directly targets to reduce the sum MSE and thus outperforms all the other distributed algorithms after few iterations. The proposed distributed precoding designs eventually provide a higher sum-group rate than the \textit{Centralized}. In fact, the iterative bi-directional training involves multiple uplink-downlink training instances with independent AWGN realizations, whereas only one (antenna-specific) noisy channel estimate is used in the \textit{Centralized} (see~\cite{Atz21}). Therefore, the impact of AWGN on the distributed precoding designs is averaged out over the iterations and, eventually, the \textit{Distributed~BR} and \textit{Distributed~GB} outperform the \textit{Centralized}. As expected, the \textit{Local~MMSE} is the best among the local precoding designs as it exploits the local interference covariance matrix that is not considered in the \textit{Local~MF}.

\begin{figure}[t]
\centering
\begin{tikzpicture}

\begin{axis}[
	width=8cm,
	height=6.5cm,
	xmin=1, xmax=30,
	ymin=0, ymax=65,
    xlabel={Bi-directional training iterations},
    ylabel={Average effective sum-group rate [bps/Hz]},
	xtick={1,5,10,15,20,25,30},
    ytick={0,10,20,30,40,50,60},
    xlabel near ticks,
	ylabel near ticks,
    x label style={font=\footnotesize},
	y label style={font=\footnotesize},
    ticklabel style={font=\footnotesize},
    legend style={at={(0.98,0.98)}, anchor=north east},
	legend style={font=\scriptsize, inner sep=1pt, fill opacity=0.75, draw opacity=1, text opacity=1},
	legend cell align=left,
	grid=both,
]

\addplot[thick, green, mark=triangle, dashed, mark options={solid}]
table[x=xin, y=locMMSE, col sep=comma] 
{Figdata/fig4data.txt};
\addlegendentry{Local MMSE};

\addplot[thick, red, mark=diamond, dashed, mark options={solid}]
table[x=xin, y=locMF, col sep=comma] 
{Figdata/fig4data.txt};
\addlegendentry{Local MF};

\addplot[thick, magenta, mark=star, mark options={solid}]
table[x=xin, y=disBR, col sep=comma] 
{Figdata/fig4data.txt};
\addlegendentry{Distributed BR};

\addplot[thick, cyan,mark=o, mark options={solid}]
table[x=xin, y=disBR-GS, col sep=comma] 
{Figdata/fig4data.txt};
\addlegendentry{Distributed BR-GS};

\addplot[thick, blue, mark=x]
table[x=xin, y=disGB, col sep=comma] 
{Figdata/fig4data.txt};
\addlegendentry{Distributed GB};

\addplot[green, mark=*, only marks, mark options={solid, scale=1.5}]
table[x=xin, y=locMMSE, col sep=comma, restrict x to domain=3:3] 
{Figdata/fig4data.txt};

\addplot[red, mark=*, only marks, mark options={solid, scale=1.5}]
table[x=xin, y=locMF, col sep=comma, restrict x to domain=6:6] 
{Figdata/fig4data.txt};

\addplot[magenta, mark=*, only marks, mark options={solid, scale=1.5}]
table[x=xin, y=disBR, col sep=comma, restrict x to domain=4:4] 
{Figdata/fig4data.txt};

\addplot[cyan, mark=*, only marks, mark options={solid, scale=1.5}]
table[x=xin, y=disBR-GS, col sep=comma, restrict x to domain=6:6] 
{Figdata/fig4data.txt};

\addplot[blue, mark=*, only marks, mark options={solid, scale=1.5}]
table[x=xin, y=disGB, col sep=comma, restrict x to domain=7:7] 
{Figdata/fig4data.txt};

\end{axis}

\end{tikzpicture}
\caption{Average effective sum-group rate versus number of bi-directional training iterations, with $r_{\mathrm{t}} = 1000$.}
\label{fig:effrateVsItr}
\end{figure}
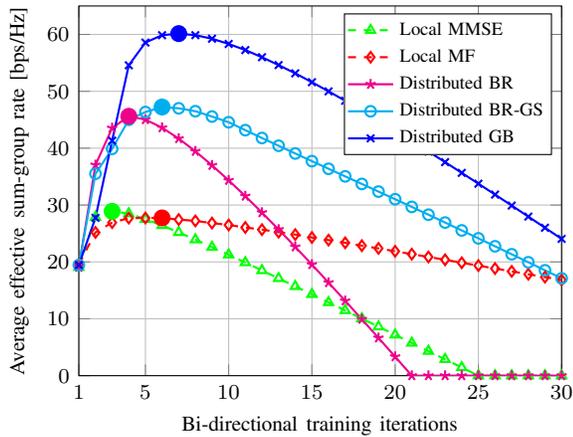

In the following, we compare the effective performance of the distributed precoding designs in terms of effective sum-group rate, defined as
\begin{align}
R_{\mathrm{eff}}^{(i)} \triangleq \bigg(1 - i \frac{r_{\mathrm{ce}}}{r_\mathrm{t}} \bigg) R^{(i)},
\end{align}
where $r_{\mathrm{ce}}$ is the number of pilot symbols used in each bi-directional training iteration and $r_{\mathrm{t}}$ is the resource block size including the transmission of both pilot symbols and data symbols. The switching time between uplink and downlink training instances is neglected. Figure~\ref{fig:effrateVsItr} plots the average effective sum-group rate as a function of the number of bi-directional training iterations with resource block size $r_{\mathrm{t}} = 1000$. All the algorithms achieve the maximum effective sum-group rate (indicated by the larger dots) within few iterations. After the peak, the performance starts to decrease as the number of data symbols transmitted within the resource block reduces at each bi-directional training iteration. As shown in Table~\ref{tab:pilots}, the \textit{Distributed~BR-GS} uses fewer pilot symbols than the \textit{Distributed~BR}. As a result, the effective sum-group rate of the \textit{Distributed~BR-GS} is slightly higher than that of the \textit{Distributed~BR}. Note that the sum-group rate (which does not consider the training overhead) of the \textit{Distributed~BR-GS} is inferior to that of the \textit{Distributed~BR} (as shown in Figure~\ref{fig:rateVsItr}). Similarly, the performance of the \textit{Local~MF} is close to that of the \textit{Local~MMSE} because fewer pilot symbols are required per bi-directional training iteration. The effective sum-group rate of the \textit{Distributed~GB} is superior to those of all the other methods. Furthermore, its training overhead is smaller than in the \textit{Distributed~BR} due to the use of group-specific pilots. In this example, the maximum effective sum-group rates of the \textit{Distributed~BR}, the \textit{Distributed~BR-GS}, and the \textit{Distributed~GB} are $1.6$, $1.65$, and $2.1$ times higher, respectively, than that of the local precoding designs.

Figure~\ref{fig:effrateVsCB_G8_K32} depicts the average effective sum-group rate as a function of the resource block size $r_{\mathrm{t}}$. For $r_{\mathrm{t}} = 1000$, the effective sum-group rates correspond to the maximum values in Figure~\ref{fig:effrateVsItr}. Note that the optimal number of bi-directional training iterations to obtain the maximum effective sum-group rate increases with $r_{\mathrm{t}}$ as a higher training overhead can be tolerated for larger resource blocks. In general, the distributed precoding designs perform well for $r_{\mathrm{t}} \geq 500$. For example, with $r_{\mathrm{t}} = 500$, the \textit{Distributed~GB} greatly outperforms all the other methods. Furthermore, the \textit{Distributed~BR-GS} performs better than the \textit{Distributed~BR} due to the use of fewer pilot symbols in each bi-directional training iteration and despite the extra interference term in \eqref{eq:diGSPreSimp}. With large resource blocks, the training overhead becomes insignificant and the effective sum-group rate approaches the sum-group rate in Figure~\ref{fig:rateVsItr}, which does not account for the training overhead.

\begin{figure}[t]
\centering
\begin{tikzpicture}

\begin{axis}[
	width=8cm,
	height=6.5cm,
	xmin=100, xmax=5000,
	ymin=10, ymax=75,
    xlabel={$r_\mathrm{t}$},
    ylabel={Average effective sum-group rate [bps/Hz]},
	xtick={100,1000,2000,3000,4000,5000},
    ytick={10,20,30,40,50,60,70,80},
	xticklabels={100,1000,2000,3000,4000,5000},
    xlabel near ticks,
	ylabel near ticks,
    x label style={font=\footnotesize},
	y label style={font=\footnotesize},
    ticklabel style={font=\footnotesize},
    legend style={at={(0.98,0.02)}, anchor=south east},
	legend style={font=\scriptsize, inner sep=1pt, fill opacity=0.75, draw opacity=1, text opacity=1},
	legend cell align=left,
	grid=both,
]

\addplot[thick, green,mark=triangle, dashed, mark options={solid}]
table[x=xin, y=locMMSE, col sep=comma] 
{Figdata/fig5data.txt};
\addlegendentry{Local MMSE};

\addplot[thick,  red, mark=diamond, dashed, mark options={solid}]
table[x=xin, y=locMF, col sep=comma] 
{Figdata/fig5data.txt};
\addlegendentry{Local MF};

\addplot[thick, magenta, mark=star, mark options={solid}]
table[x=xin, y=disBR, col sep=comma] 
{Figdata/fig5data.txt};
\addlegendentry{Distributed BR};

\addplot[thick, cyan,mark=o, mark options={solid}]
table[x=xin, y=disBR-GS, col sep=comma] 
{Figdata/fig5data.txt};
\addlegendentry{Distributed BR-GS};

\addplot[thick, blue, mark=x]
table[x=xin, y=disGB, col sep=comma] 
{Figdata/fig5data.txt};
\addlegendentry{Distributed GB};

\end{axis}

\end{tikzpicture}
\caption{Average effective sum-group rate versus resource block size.}
\label{fig:effrateVsCB_G8_K32}
\end{figure}
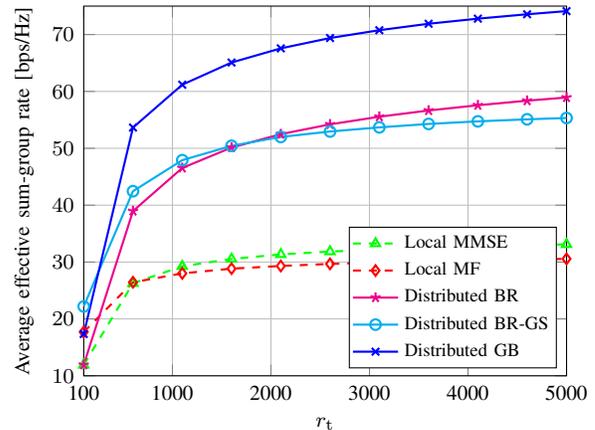

\begin{figure}[t]
\centering
\begin{tikzpicture}

\begin{axis}[
	width=8cm,
	height=6.5cm,
	xmin=2, xmax=16,
	ymin=5, ymax=70,
    xlabel={$|\setK_g|$},
    ylabel={Average effective sum-group rate [bps/Hz]},
	xtick={2,4,8,16},
    ytick={10,20,30,40,50,60,70},
    xlabel near ticks,
	ylabel near ticks,
    x label style={font=\footnotesize},
	y label style={font=\footnotesize},
    ticklabel style={font=\footnotesize},
    legend style={at={(0.02,0.02)}, anchor=south west},
	legend style={font=\scriptsize, inner sep=1pt, fill opacity=0.75, draw opacity=1, text opacity=1},
	legend cell align=left,
	grid=both,
    xmode=log,
    log basis x={2}
]

\addplot[thick, green,mark=triangle, dashed, mark options={solid}]
table[x=xin, y=locMMSE, col sep=comma] 
{Figdata/fig6data.txt};
\addlegendentry{Local MMSE};

\addplot[thick,  red, mark=diamond, dashed, mark options={solid}]
table[x=xin, y=locMF, col sep=comma] 
{Figdata/fig6data.txt};
\addlegendentry{Local MF};

\addplot[thick, magenta, mark=star, mark options={solid}]
table[x=xin, y=disBR, col sep=comma] 
{Figdata/fig6data.txt};
\addlegendentry{Distributed BR};

\addplot[thick, cyan,mark=o, mark options={solid}]
table[x=xin, y=disBR-GS, col sep=comma] 
{Figdata/fig6data.txt};
\addlegendentry{Distributed BR-GS};

\addplot[thick, blue, mark=x]
table[x=xin, y=disGB, col sep=comma] 
{Figdata/fig6data.txt};
\addlegendentry{Distributed GB};

\end{axis}

\end{tikzpicture}
\caption{Average effective sum-group rate versus number of UEs in each multicast group, with $r_{\mathrm{t}} = 1000$.}
\label{fig:effrateVsCB_G8_K64}
\end{figure}
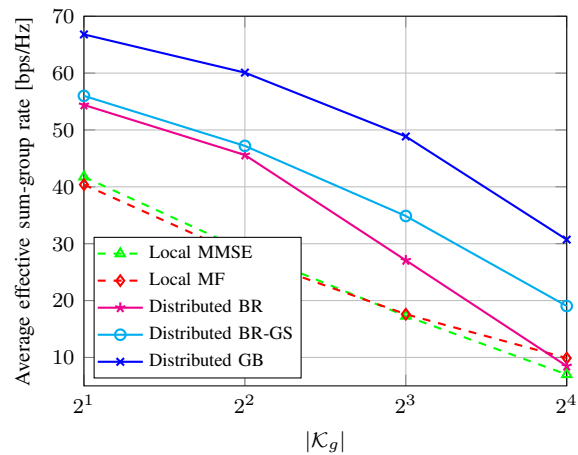

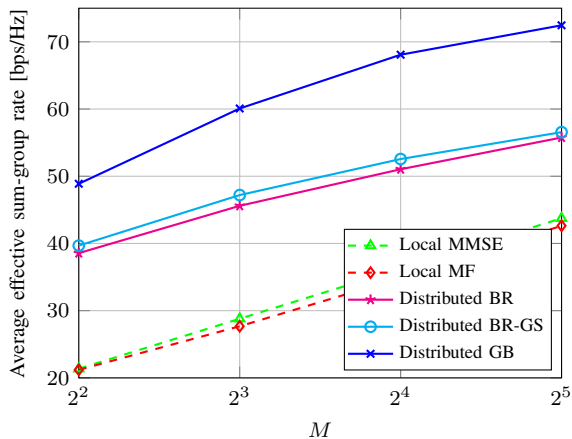
\begin{figure}[t]
\centering
\begin{tikzpicture}

\begin{axis}[
	width=8cm,
	height=6.5cm,
	xmin=4, xmax=32,
	ymin=20, ymax=75,
    xlabel={$M$},
    ylabel={Average effective sum-group rate [bps/Hz]},
	xtick={4,8,16,32},
    ytick={20,30,40,50,60,70},
    xlabel near ticks,
	ylabel near ticks,
    x label style={font=\footnotesize},
	y label style={font=\footnotesize},
    ticklabel style={font=\footnotesize},
    legend style={at={(0.98,0.02)}, anchor=south east},
	legend style={font=\scriptsize, inner sep=1pt, fill opacity=0.75, draw opacity=1, text opacity=1},
	legend cell align=left,
	grid=both,
    xmode=log,
    log basis x={2}
]

\addplot[thick, green,mark=triangle, dashed, mark options={solid}]
table[x=xin, y=locMMSE, col sep=comma] 
{Figdata/fig7data.txt};
\addlegendentry{Local MMSE};

\addplot[thick,  red, mark=diamond, dashed, mark options={solid}]
table[x=xin, y=locMF, col sep=comma] 
{Figdata/fig7data.txt};
\addlegendentry{Local MF};

\addplot[thick, magenta, mark=star, mark options={solid}]
table[x=xin, y=disBR, col sep=comma] 
{Figdata/fig7data.txt};
\addlegendentry{Distributed BR};

\addplot[thick, cyan,mark=o, mark options={solid}]
table[x=xin, y=disBR-GS, col sep=comma] 
{Figdata/fig7data.txt};
\addlegendentry{Distributed BR-GS};

\addplot[thick, blue, mark=x]
table[x=xin, y=disGB, col sep=comma] 
{Figdata/fig7data.txt};
\addlegendentry{Distributed GB};

\end{axis}

\end{tikzpicture}
\caption{Average effective sum-group rate versus number of antennas at each BS, with $r_{\mathrm{t}} = 1000$.}
\label{fig:effrateVsCB_M32}
\end{figure}

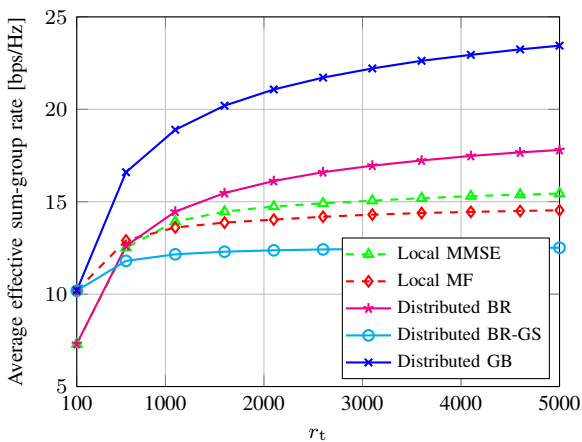
\begin{figure}[t]
\centering
\begin{tikzpicture}

\begin{axis}[
	width=8cm,
	height=6.5cm,
	xmin=100, xmax=5000,
	ymin=5, ymax=25,
    xlabel={$r_\mathrm{t}$},
    ylabel={Average effective  sum-group rate [bps/Hz]},
    xtick={100,1000,2000,3000,4000,5000},
    ytick={5,10,15,20,25},
	xticklabels={100,1000,2000,3000,4000,5000},
    xlabel near ticks,
	ylabel near ticks,
    x label style={font=\footnotesize},
	y label style={font=\footnotesize},
    ticklabel style={font=\footnotesize},
    legend style={at={(0.98,0.02)}, anchor=south east},
	legend style={font=\scriptsize, inner sep=1pt, fill opacity=0.75, draw opacity=1, text opacity=1},
	legend cell align=left,
	grid=both,
]

\addplot[thick, green,mark=triangle, dashed, mark options={solid}]
table[x=xin, y=locMMSE, col sep=comma] 
{Figdata/fig8data.txt};
\addlegendentry{Local MMSE};

\addplot[thick,  red, mark=diamond, dashed, mark options={solid}]
table[x=xin, y=locMF, col sep=comma] 
{Figdata/fig8data.txt};
\addlegendentry{Local MF};

\addplot[thick, magenta, mark=star, mark options={solid}]
table[x=xin, y=disBR, col sep=comma] 
{Figdata/fig8data.txt};
\addlegendentry{Distributed BR};

\addplot[thick, cyan,mark=o, mark options={solid}]
table[x=xin, y=disBR-GS, col sep=comma] 
{Figdata/fig8data.txt};
\addlegendentry{Distributed BR-GS};

\addplot[thick, blue, mark=x]
table[x=xin, y=disGB, col sep=comma] 
{Figdata/fig8data.txt};
\addlegendentry{Distributed GB};

\end{axis}

\end{tikzpicture}
\caption{Average effective sum-group rate versus resource block size at low SNR ($\sigma^2_{\bs}=\sigma^2_{\ue} =-75$~dBm).}
\label{fig:effrateVsCB_sigma_75dBm}
\end{figure}

Figure~\ref{fig:effrateVsCB_G8_K64} plots the average effective sum-group rate as a function of the number of UEs in each multicast group $|\setK_g|$ with resource block size $r_{\mathrm{t}} = 1000$. In general, the sum-group rate decreases when $|\setK_g|$ grows as more spatial degrees of freedom are used to suppress the interference among the multicast groups. However, at the same time, the sum rate across all the UEs $\sum_{g \in \setG} |\setK_g| R_g$ is increased. The training overhead associated with $\X_k^{\ulA}$ in the \textit{Distributed~BR} and the \textit{Local~MMSE} depends on $K=\sum_{g \in \setG} |\setK_g|$, while the training overhead associated with $\X_k^{\ulB}$ in the \textit{Distributed~BR-GS}, the \textit{Distributed~GB}, and the \textit{Local~MF} is dictated by $G$. Consequently, the \textit{Distributed~BR} and the \textit{Local~MMSE} are more severely penalized by an increase in $K$. For example, considering the case of $|\setK_g|=16$, the performance of the \textit{Distributed~BR} and the \textit{Local~MMSE} is inferior even to that of the \textit{Local~MF}.

Figure~\ref{fig:effrateVsCB_M32} depicts the average effective sum-group as a function of the number of antennas at each BS with resource block size $r_{\mathrm{t}} = 1000$. Increasing $M$ obviously improves the performance of all the considered methods. What is more, the proposed distributed methods provide significant gains over the local precoding designs even with a relatively high number of antennas at each BS, e.g., $M=32$, which motivates the use of the distributed precoding designs even in such scenarios.

Figure~\ref{fig:effrateVsCB_sigma_75dBm} illustrates the average effective sum-group as a function of the resource block size at low SNR, where the joint interference suppression across the BSs becomes less important. Nonetheless, the \textit{Distributed~GB} is superior to all the other methods, whereas the \textit{Distributed~BR-GS}, which depends now on the noisy feedback with the extra interference term in \eqref{eq:diGSPreSimp}, suffers from inaccuracies in the local interference covariance matrix, making it inferior to the local precoding designs.

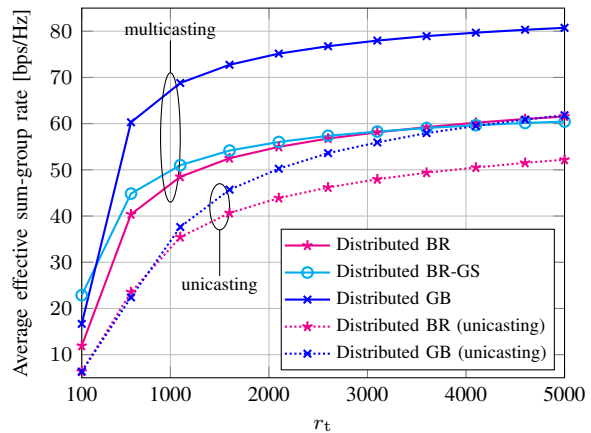
\begin{figure}[t]
\centering
\begin{tikzpicture}

\begin{axis}[
	width=8cm,
	height=6.5cm,
	xmin=100, xmax=5000,
	ymin=5, ymax=85,
    xlabel={$r_\mathrm{t}$},
    ylabel={Average effective sum-group rate [bps/Hz]},
	xtick={100,1000,2000,3000,4000,5000},
    ytick={0,10,20,30,40,50,60,70,80},
	xticklabels={100,1000,2000,3000,4000,5000},
    xlabel near ticks,
	ylabel near ticks,
    x label style={font=\footnotesize},
	y label style={font=\footnotesize},
    ticklabel style={font=\footnotesize},
    legend style={at={(0.98,0.02)}, anchor=south east},
	legend style={font=\scriptsize, inner sep=1pt, fill opacity=0.75, draw opacity=1, text opacity=1},
	legend cell align=left,
	grid=both,
]

\addplot[thick, magenta, mark=star, mark options={solid}]
table[x=xin, y=disBR, col sep=comma] 
{Figdata/fig9data.txt};
\addlegendentry{Distributed BR};

\addplot[thick, cyan,mark=o, mark options={solid}]
table[x=xin, y=disBR-GS, col sep=comma] 
{Figdata/fig9data.txt};
\addlegendentry{Distributed BR-GS};

\addplot[thick, blue, mark=x]
table[x=xin, y=disGB, col sep=comma] 
{Figdata/fig9data.txt};
\addlegendentry{Distributed GB};

\addplot[thick, magenta, densely dotted, mark=star, mark options={solid}]
table[x=xin, y=disBR_uni, col sep=comma] 
{Figdata/fig9data.txt};
\addlegendentry{Distributed BR (unicasting)};

\addplot[thick, blue, densely dotted,  mark=x, mark options={solid}]
table[x=xin, y=disGB_uni, col sep=comma] 
{Figdata/fig9data.txt};
\addlegendentry{Distributed GB (unicasting)};

\draw (1000,57) ellipse (100 and 14);
\draw (1500,42) ellipse (100 and 5);

\begin{scope}[>=latex]
\draw[-] (1000,71) -- (1000,78) {};
\draw[-] (1500,37) -- (1500,27) {};
\end{scope}

\node[black, font=\scriptsize] at (1000,80) {multicasting};
\node[black, font=\scriptsize] at (1500,25) {unicasting};

\end{axis}

\end{tikzpicture}
\caption{Average effective sum-group rate versus resource block size in comparison with the unicast precoding design~\cite{Atz21} with the same CSI accuracy.}
\label{fig:mc_vs_uc}
\end{figure}

Lastly, Figure~\ref{fig:mc_vs_uc} compares the proposed distributed multi-group multicast precoding designs with the distributed unicast precoding design developed in~\cite{Atz21} in the multi-group multicasting scenario considered so far (i.e., with $K=32$~UEs divided into $G=8$~multicast groups of $4$~randomly selected UEs). The unicast precoding design is intended to suppress the interference among all the UEs and does not consider that the latter are divided into groups. For this method, the same data symbols (distinctly modulated for each UE) are transmitted to all the UEs in a multicast group by means of UE-specific precoders. In this setting, the rate is still limited by the worst UE in the multicast group and, therefore, we use the sum-group rate in \eqref{eq:R} as a metric to evaluate the performance of the unicast precoding design. Moreover, the unicast precoding design requires UE-specific pilots, which are longer than the group-specific pilots used for the multicast precoding designs and thus result in higher CSI accuracy. Hence, to compare the impact of the training overhead between the multicast and unicast precoding designs, we scale the transmit power of the group-specific pilots to achieve the same CSI accuracy as the UE-specific pilots. Figure~\ref{fig:mc_vs_uc} plots the average effective sum-group rate as a function of the resource block size. Here, the \textit{Distributed~BR (unicasting)} indicates the distributed precoding design proposed in~\cite{Atz21} while \textit{Distributed~GB (unicasting)} corresponds to Algorithm~\ref{alg:disGd} adapted to consider each UE as a multicast group. We observe that all the proposed distributed methods tailored for the multi-group multicasting scenario outperform the unicast precoding designs. In addition, with small resource blocks, the performance of the unicast precoding designs is further penalized due to the higher impact of the training overhead. For instance, with $r_{\mathrm{t}} = 1000$, the \textit{Distributed~GB (unicasting)} delivers around $4.5$~bps/Hz per UE, while the \textit{Distributed~GB} provides approximately $8.5$~bps/Hz per UE. We point out that even the performance of the multicast precoding designs with non-scaled transmit power of the group-specific pilots in Figure~\ref{fig:effrateVsCB_G8_K32} is significantly better than that of the unicast precoding designs in Figure~\ref{fig:mc_vs_uc}.

\section{Conclusions} \label{sec:CONC}

We proposed fully distributed multi-group multicast precoding designs for cell-free massive MIMO systems with modest training overhead. The sum-group MSE minimization is initially considered to guarantee absolute MSE fairness within each multicast group. Subsequently, to simplify the computation and signaling, the sum-group MSE is approximated with the sum MSE objective. Considering the UE-specific rates as the performance metric, the aforementioned approximation holds well, especially at high SNR. An iterative bi-directional training is adopted to design the precoders and the combiners locally at each BS and at each UE, respectively. To this end, a new group-specific OTA uplink training resource is introduced to obtain the required group-specific cross terms from other BSs in the distributed precoding design, which eliminates the need for backhaul signaling to exchange the CSI. Furthermore, the distributed precoding designs are implemented by means of either best-response or gradient-based updates exploiting UE- and/or group-specific pilots. Consequently, the distributed precoding design with best-response updates results in a steepest descent direction for the sum MSE minimization, which makes it inferior to its centralized implementation. However, the gradient-based update solves the sum MSE minimization as it would be in a centralized design. Numerical results show that the distributed gradient-based precoding design with group-specific pilots always yields the best effective performance. Moreover, all the proposed distributed methods greatly outperform conventional cell-free massive MIMO precoding designs that rely solely on local CSI.

\appendices

\section{Sub-Gradient Update of $\{\mu_k\}_{k \in \setK}$ and $\{\lambda_b\}_{b \in \setB}$} \label{app:A1}

\textit{1) Sub-gradient update of $\{\mu_k\}_{k \in \setK}$.} The optimality condition for $t_g$ is given by 
\begin{align}\label{eq:nuCon}
\frac{\partial}{\partial t_g} \mathcal{L}_{\eqref{eq:probBS}} \big( \{ \w_g, t_g, \mu_{k}, \lambda_b \} \big) = 0 \implies \sum_{k \in \setK_g} \mu_{k} = 1.
\end{align}
Moreover, the complementary slackness conditions corresponding to the per-UE MSE constraint in \eqref{eq:probFor} are given by
\begin{align} \label{eq:comSl2}
& \mu_{k} (\mse_{k} - t_{g}) = 0, \quad \forall k \in \setK_g \nonumber \\
& \implies \sum_{k \in \setK_g} \mu_{k} t_{g} = \sum_{k \in \setK_g} \mu_{k}\mse_{k}.
\end{align}
Therefore, from \eqref{eq:nuCon} and \eqref{eq:comSl2}, we have $t_g = \sum_{k \in \setK_g} \mu_{k} \mse_{k}$. To achieve absolute MSE fairness within each multicast group, each $\mu_k$ is updated as~\cite{Mah21}
\begin{align} \label{eq:nusgud}
\mu^{(i)}_k & = \max \bigg( 0, \mu^{(i-1)}_k + \zeta \frac{\partial}{\partial \mu_k} \mathcal{L}_{\eqref{eq:probBS}} \big( \{ \w_g, t_g, \mu_{k}, \lambda_b \} \big) \bigg) \\
& = \max \bigg( 0, \mu^{(i-1)}_k + \zeta (\mse_{k} - t_{g}) \bigg),
\end{align}
where $i$ is the iteration index and $\zeta$ is the step size. Finally, \eqref{eq:nusgud} is normalized to meet the constraint in \eqref{eq:nuCon}.

\smallskip

\textit{2) Sub-gradient update of $\{\lambda_b\}_{b \in \setB}$.} To meet the per-BS transmit power constraint, $\lambda_b$ is updated as~\cite{Kom13}
\begin{align} \label{eq:lambdab}
\lambda^{(i)}_b  & = \max \bigg( 0, \lambda^{(i-1)}_b + \eta \frac{\partial}{\partial \lambda_b} \mathcal{L}_{\eqref{eq:probBS}} \big( \{ \w_g, t_g, \mu_{k}, \lambda_b \} \big) \bigg) \\
& = \max \bigg( 0, \lambda^{(i-1)}_b + \eta \bigg( \sum_{g \in \setG} \| \w_{b,g} \|^2 - \rho_{\bs} \bigg) \bigg),
\end{align}
where $\eta$ is the step size.

\section{KKT Conditions of \eqref{eq:appxsmse}} \label{app:A2}

The Lagrangian of \eqref{eq:appxsmse} can be written as
\begin{equation}
\mathcal{L}_{\eqref{eq:appxsmse}} \big( \{ p_g, \kappa \} \big) \triangleq \sum_{k \in \setK} \frac{\sigma^2_{\ue}}{p_{g_k} c_{kk}^2} + \kappa \ \bigg( \sum_{\bar g \in \setG} {p_{\bar g}} - \rho_{\bs} \bigg),
\end{equation}
where $\kappa$ is the dual variable corresponding to the constraint in \eqref{eq:appxsmse}. Then, the optimal $p_g$ is obtained as
\begin{equation}
\frac{\partial}{\partial{p_g}} \mathcal{L}_{\eqref{eq:appxsmse}}\big(\{p_g, \kappa \}\big)=0 \implies p_g = \sqrt{\sum_{k \in \setK_g} \frac{\sigma^2_{\ue}}{\kappa c_{kk}^2}}.
\end{equation}
Finally, $\kappa$ is computed to satisfy $\sum_{g \in \setG} p_g = \rho_{\bs}$, which yields
\begin{equation}
\kappa = \frac{1}{\rho^2_{\bs}} \bigg( \sum_{g \in \setG} \sqrt{\sum_{k \in \setK_g} \frac{\sigma^2_{\ue}}{c_{kk}^2}} \bigg)^2.
\end{equation}

\section{Local Precoding Designs} \label{app:A3}

To avoid the prohibitive complexity and backhaul signaling of large-scale centralized precoding designs, most works on cell-free massive MIMO assume simple local precoding strategies exploiting the large-antenna regime across the BSs~\cite{Chen21}. In this setting, the BS-specific precoders are designed based solely on local CSI, ignoring the contribution from the other BSs. Nevertheless, iterative bi-directional training is required to update the precoders at the BSs based on the combiners at the UEs and vice versa. With perfect CSI, at each bi-directional training iteration, the \textit{Local~MMSE} precoder at each BS~$b$ is computed as
\begin{align}\label{eq:localMSE}
\w_{b,g} = \bigg( \sum_{k \in \setK} \omega_{k} \H_{b, k} \v_{ k} \v_{ k}^{\herm} \H_{b, k}^{\herm} \! + \! \lambda_{b}\I_M \bigg)^{-1} \sum_{k \in \setK_g} \omega_{k} \H_{b,k} \v_{k},
\end{align}
whereas the corresponding \textit{Local~MF} precoder is computed as
\begin{align}\label{eq:CBproj}
\w_{b,g} = \frac{1}{\lambda_b} \sum_{ k\in \setK_g}\omega_{k} \H_{b,k} \v_{k}.
\end{align}
Note that the dual variable $\lambda_{b}$ in \eqref{eq:localMSE} and \eqref{eq:CBproj} can be easily obtained via bisection. In both cases, each UE~$k$ computes its combiner as in \eqref{eq:uebf}. The local precoding designs may not convergence to a solution of the sum MSE minimization in \eqref{eq:EqvprobForHi}. However, the resulting UE-specific rates improve over the iterations since the combiners are better focused towards the intended signals and increase the accuracy of the effective channel estimation. The practical implementation of the \textit{Local~MMSE} and the \textit{Local~MF} requires, at each bi-directional training iteration, the UE- and group-specific effective uplink channel estimations, respectively, as well as the effective downlink channel estimation (see Section~\ref{sec:SM_est}). Accordingly, the \textit{Local~MMSE} precoder at each BS~$b$ is computed as
\begin{align}
\label{eq:localMMSE} \w_{b,g} & = \sqrt{\beta^{\ulA}} \big(\Y_{b}^{\ulA} \Omegab ({\Y_{b}^{\ulA}})^{\herm} + \tau^{\ulA}({\beta^{\ulA}}\lambda_{b} - \sigma_{\bs}^2)\I_M\big)^{-1} \nonumber \\ 
& \phantom{=} \ \times \sum_{k \in \setK_g} \omega_k\Y_{b}^{\ulA}\p^{\ulA}_k,
\end{align}
with $\Omegab \triangleq \Diag(\omega_1, \ldots, \omega_K) \in \Real^{K \times K}$, whereas the corresponding \textit{Local~MF} precoder is computed as
\begin{align}
\label{eq:conjBf} \w_{b,g} = \frac{1}{\lambda_b\tau^{\ulB} \sqrt{\beta^{\ulB}}} \Y_{b}^{\ulB}\p^{\ulB}_g.
\end{align}
In both cases, each UE~$k$ computes its combiner as in \eqref{eq:rxmmse}. If pilot contamination is to be avoided entirely, the \textit{Local~MMSE} requires a minimum of $K \geq G$ orthogonal pilots to obtain $\Y_{b}^{\ulA}$ in \eqref{eq:Y_b_ul1} in each uplink training instance, whereas the \textit{Local~MF} requires a minimum of $G$ orthogonal pilots to obtain $\Y_{b}^{\ulB}$ in \eqref{eq:Y_b_ul2} in each uplink training instance. For a fixed number of bi-directional iterations, the \textit{Local~MMSE} outperforms the \textit{Local~MF} by exploiting the local interference covariance matrix, although it has a higher training overhead.

\bibliographystyle{IEEEtran}
\bibliography{IEEEabbr,refs}

\end{document}